\documentclass[conference]{IEEEtran}

\usepackage{graphicx} 
\usepackage[hyphens]{url}
\usepackage{hyperref}
\usepackage{array,multirow}
\usepackage{booktabs}
\usepackage[footnotesize]{caption} 
\usepackage{tabularx}
\usepackage{ctable}
\usepackage{amsmath,amsfonts,amssymb}
\usepackage{amsthm}

\usepackage{tikz}
\usepackage{tkz-base}
\usepackage{pgf-pie}
\usepackage{caption}
\usepackage{subcaption}

\begin{document}

\title{\emph{MyTrackingChoices}: Pacifying the Ad-Block War by Enforcing User Privacy Preferences}

\author{\IEEEauthorblockN{Jagdish Prasad Achara}
\IEEEauthorblockA{INRIA Grenoble\\
France\\
Email: jagdish.achara@inria.fr}
\and
\IEEEauthorblockN{Javier Parra-Arnau}
\IEEEauthorblockA{INRIA Grenoble\\
France\\
Email: xparnau@gmail.com}
\and
\IEEEauthorblockN{Claude Castelluccia}
\IEEEauthorblockA{INRIA Grenoble\\
France\\
Email: claude.castelluccia@inria.fr}}

\maketitle
\begin{abstract}
Free content and services on the Web are often supported by ads.
However, with the proliferation of intrusive and privacy-invasive ads, a significant proportion of users have started to use ad blockers.
As existing ad blockers are radical (they block all ads) and are not designed taking into account their economic impact, ad-based economic model of the Web is in danger today.

In this paper, we target privacy-sensitive users and provide them with fine-grained control over tracking.
Our working assumption is that some categories of web pages (for example, related to health, religion, etc.) are more privacy-sensitive to users than others (education, science, etc.).
Therefore, our proposed approach consists in providing users with an option to specify the categories of web pages that are privacy-sensitive to them and block trackers present on such web pages only.
As tracking is prevented by blocking network connections of third-party domains, we avoid not only tracking but also third-party ads.
Since users will continue receiving ads on web pages belonging to non-sensitive categories, our approach essentially provides a trade-off between privacy and economy.
To test the viability of our solution, we implemented it as a Google Chrome extension, named \emph{MyTrackingChoices} (available on Chrome Web Store).
Our real-world experiments with \emph{MyTrackingChoices} show that the economic impact of ad blocking exerted by privacy-sensitive users can be significantly reduced.

\end{abstract}


\section{Introduction}
\label{sec:intro}

According to PageFair 2015 report~\cite{pagefaire_report_2015}, 45 million Americans (16\% of online users) and 77 million Europeans had installed ad blockers
\footnote{All tools 
that eventually end up blocking ads even though they are initially designed for some other purpose (privacy, transparency, etc.)
} 
as of the second quarter of 2015.
All this globally accounts for 21.8 billion dollars worth of blocked ads in the year 2015.
Since ads fuel the free content and services over the Web, this monetary loss puts the ad-based economic model of the Web under threat.
The nature of this threat has become more devastating over time since,
in the early days of online advertising, ad industry ignored this potentially dangerous situation in favor of immediate benefits.

Users generally block ads due to an unpleasant ad-experience, which could be a fallout of several concerns. 
Some users simply do not want to see ads, while others block ads because they find them too annoying (in terms of usability) or perceive them as a source of privacy and/or security issues on the Web~\cite{pagefaire_report_2015,targeted_ads_harmful,a_way_to_peace,profiling_adblockers,dcn_adblock_report,why_millennials_block_ads,why_people_block_ads,survey_search_engine_use,americans_reject_survey}.
In fact, the reasons behind ad blocking are not mutually exclusive and some users may block ads as a result of their combinations.
Apparently, the general ad-experience has deteriorated over the last few years, and this has resulted into an unprecedented use of ad blockers on the Web.

Apart from the poor ad-experience, radical design choices adopted 
by ad blockers is another reason why ad blocking has become a threat 
to the Web economy today. A scrutiny of current ad blockers further 
reveals that they do not take into account: 1) the economic impact 
of ad blocking, and 2) the social and economic benefits of non-intrusive 
and rational advertising. This demands better tools that cater to users' 
concerns, while at the same time are designed to ensure that their economic 
impact does not get sidelined.

In this work, we address the evoked concerns by proposing a three-dimensional approach that combines user privacy, ad blocking and its economic impact.
We target users who are not against ads but block them due to privacy concerns.
Our working assumption is that users consider their visits to some web pages (for example, related to religion, health) more privacy-sensitive than to others (such as news, sports)~\cite{Leon_2013,chanchary2015user}.
Therefore, our approach consists in providing users an option to specify the categories of web pages where they do not want to be tracked and thus, do not want to receive third-party ads\footnote{Such ads are delivered by domains other than the one navigated by a user.}.
Unlike current ad blockers, it allows users to continue receiving ads on categories of web pages that they consider non-sensitive.
In concrete terms, the proposed approach attempts to find a trade-off between privacy and economy.

To test the viability of our approach, we have developed an extension for Google Chrome called \emph{MyTrackingChoices}\footnote{\url{https://chrome.google.com/webstore/detail/mytrackingchoices/fmonkjimgifgcgeocdhhgbfoncmjclka?hl=fr}}. 
While users browse the web, the extension categorizes the visited web page on the fly and, 
depending on the users' choices, it blocks the network connections of third-party domains present on the web page.
As tracking is prevented by blocking third-party network requests, we avoid not only tracking but also third-party ads.
It is worth noting that, unlike other ad blockers, our extension does not block ads served directly from the publisher on both sensitive and non-sensitive web pages.

\textbf{Contributions.} We make the following contributions in this paper.
\begin{itemize}
\item We review existing ad blockers for their design choices and the ensuing impact on the Web economy. Our findings reveal that their approach to ad blocking is not appropriate both to users and the economy. We also show that current self-regulatory initiatives by the ad industry and various other organizations are not sufficient because users' choices are not guaranteed to be enforced.

\item Building on the lessons learned, we propose a new approach that is user-centric and gives fine-grained control to users.
Being user-centric ensures that user privacy and ad preferences are enforced on the user-side while fine-grained control provides users a a trade-off between privacy and economy.
Our solution consists in providing users with an option to select categories of web pages that are privacy-sensitive to them and block network connections of third-party domains (and thus, ads delivered through third-party domains) present on such web pages.
To test the viability of our proposed approach, we implement it as a Google Chrome extension.

\item We present some preliminary results based on the pseudo-anonymous data collected from 96 users of the \emph{MyTrackingChoices} extension. Our results regarding the impact of MyTrackingChoices on the Web economy show that the economic impact of ad blocking exerted by privacy-sensitive users can be significantly reduced. We also evaluate the usability and the performance aspects of MyTrackingChoices which confirm that it can be adopted on a wide scale.
\end{itemize}

\textbf{Outline.}
Section~\ref{sec:problem_adblockers} discusses the problems and perspectives associated to existing ad blockers. In Section~\ref{sec:mytrackingchoices}, we present our new approach to ad blocking and its implementation as \emph{MyTrackingChoices}. The impact of \emph{MyTrackingChoices} on the Web economy is evaluated in Section~\ref{sec:economic_impact} whereas Section~\ref{sec:usability_performance} assesses the usability and performance of \emph{MyTrackingChoices}. In the end, Section~\ref{sec:measurement_trackers} measures the presence of third-party domains and Section~\ref{sec:conclusion} concludes our paper with some directions for future work.

\section{Existing Ad Blockers:\\ Problems and Perspectives}
\label{sec:problem_adblockers}
Due to proliferation of intrusive\footnote{Ads are considered to be intrusive if 
they hide content or if they pop up randomly on the screen making user browsing experience frustrating.} 
and privacy-invasive ads, ad blockers have become very popular.
They can be divided into two classes: ``ad blockers" and ``anti-trackers".
Ad blockers are those which exists solely for blocking ads whereas anti-trackers exist for other reasons like privacy and transparency, but eventually block ads.
Some examples of ad blockers are AdBlock~\cite{Adblock16Web}, AdBlock Plus~\cite{AdblockPlus16Web}, etc. whereas Ghostery~\cite{Ghostery16Web}, Disconnect~\cite{Disconnect16Web}, PrivacyBadger~\cite{PrivacyBadger16Web} are examples of anti-trackers.
In terms of functionalities, ad blockers in the first class block all ads whereas anti-trackers allow users to block a particular tracker or a particular category (related to analytics, privacy, advertising, etc.) of trackers or trackers on a per domain basis.
Considering this distinction in terms of their functionalities and objectives, in the remainder of this section, we make the distinction between them.

\textbf{Problem with current ad blockers and anti-trackers.}
Ad blockers are too radical because they take a binary approach towards ad blocking, i.e., they block all ads.
Additionally, as they simply block all ads by default and do not let users configure their choices, it is evident that the developers have not considered their economic impact on the Web.

Anti-trackers give users the option to decide by which trackers or category of trackers they do not want to be tracked. 
However, we argue that most users are not concerned with the trackers but the other dimension of tracking, i.e., on which web pages they do not want to be tracked.
In this respect, anti-trackers let users decide to block trackers on a per domain basis, i.e., users can whitelist or blacklist a specific domain.
However, we argue that domain-level granularity is not the right approach for three reasons:
\begin{enumerate}
\item Given the huge number of domains, it is almost impossible for users to determine and pre-define all the domains where they do not want to be tracked.
\item Some domains can host web pages belonging to different categories, for example, belonging to both sensitive (health, religion, etc.) and non-sensitive (sports, science, etc.) categories~\cite{handbook_data_protection}.
Domains belonging to the ``news" category, e.g., cnn.com or bbc.com, are good examples because they usually include web pages from a variety of categories (sports, economy, politics, health, travel, religion, etc.).
Therefore, except some domains that have all web pages belonging to a same category, it makes more sense to block trackers based on the category of the web page and not on a per domain basis.
In fact, page-granular (and not domain-granular) blocking makes more sense both from users' privacy point of view and with respect to reducing the economic damage incurred by publishers.
The reason being that, for a given domain, ads would eventually be blocked only on web pages belonging to sensitive categories chosen by users and not on the whole domain.
\item Blocking trackers based on the category of web pages (instead of configuring it for each and every domain) makes it easier for users to configure as they just need to select once the categories of web pages that are sensitive to them.
As the number of web page categories are limited, these pre-defined categories can be made available to the user when the tool is installed and user choices can be respected from the very beginning.
\end{enumerate}

With the exception of PrivacyBadger~\cite{PrivacyBadger16Web}, another issue with existing ad blockers and anti-trackers is that, in order to block trackers or ads, they rely on black-lists manually maintained by their developers or, in some cases, by user communities.
The use of these lists by AdBlock Plus~\cite{AdblockPlus16Web}, currently the most popular among ad blocking tools, stirred controversy for accepting money from some ad companies to whitelist them~\cite{Cookson15FT,rip_adblock_plus}.
Furthermore, it is very cumbersome to maintain these lists as 1) new trackers and/or ad companies keep appearing in the market and 2) already existing companies keep introducing new mechanisms to deliver ads.

\textbf{Self-regulatory initiatives by the ad industry are not appropriate.}
Without giving control to the user (i.e., providing users a tool), various self-regulatory initiatives~\cite{lean, acceptableads} from the ad industry have proposed solutions to counter some of the concerns behind ad blocking (like annoyance, usability and performance).
These efforts envisage to improve the ad experience for users but none of them gives the control back to users.
Moreover, these initiatives from the ad industry do not take into account the privacy concerns behind ad blocking even though various studies confirm 
that a non-negligible number of users block ads because of privacy concerns~\cite{pagefaire_report_2015,profiling_adblockers,dcn_adblock_report}.
The LEAN program from IAB talks about non-invasive ads but it is not in terms of privacy~\cite{lean}.
``Acceptable ads manifesto" include five points to improve the ad-experience and thus, making them acceptable.
However, none of these five points tackles privacy concerns~\cite{acceptableads}.

Other examples of self-regulatory initiatives are ``Your online choices" (from a group of European organizations)~\cite{YourOnlineChoicesWeb} and DNT (from World Wide Web Consortium)~\cite{DNT15W3C}.
Your online choices lets users block ads tailored to their web browsing interests but users can never be sure if their choices are actually honored and if they are still being tracked.
These initiatives do not suffice since users' choices are not technically enforced at the user side.
Similarly, DNT allows users to notify websites and ad industry if they want to stop being tracked through third-party cookies.
However, the problem of enforcement of user choices still persists.

\begin{table*}[tb!]
\centering
\caption{Top-level interest categories.}
\label{tab:TopLevelCategories}
\footnotesize
  \begin{tabular}{cccc}
    \toprule
        adult                                & economics                       & hobbies \& interests                         &  politics \\
    agriculture                     &  education                                     &  home                                     & real estate \\
    animals                         & family \& parenting                              & law                              & religion \\
    architecture                  & fashion                                                  & military                                         & science\\
    arts \& entertainment  & folklore                                         & news                                 & society \\
    automotive                    & food \& drink                                 & personal finance                                                 & sports\\
    business                        & health \& fitness                     & pets                           & technology \& computing\\
    careers                          & history                               & philosophy                       & travel\\
    \bottomrule
  \end{tabular}
\end{table*}

\section{MyTrackingChoices}
\label{sec:mytrackingchoices}
As we have discussed in the Section~\ref{sec:problem_adblockers}, current ad blockers and existing self-regulatory initiatives are not appropriate to tackle the current threat to ad-based economic model of the Web.
Today there is a need of better tools that give users more fine-grained control over tracking and thus, third-party ads.

\subsection{New approach to ad blocking}
We propose a different approach as a solution to the afore-described concerns.
Our approach targets users who are not against ads but block them due to privacy reasons.
It is based on the assumption that most people do not want to be tracked on ``sensitive" websites (for example, related to religion, health),  
but accept to be tracked and receive ads on less sensitive ones (such as news, sport) to support the content provider.

The underlying objective is to sustain the ad-based economic model of the Web and hence, is in contrast to the design rationale of existing ad blockers.
The idea is to let users choose the categories of web pages that are privacy-sensitive to them (for example, health, religion) and exclusively block the respective trackers and/or ads on those web pages.
The granularity of selecting where users accept ads gives it an advantage over other ad blockers in various respects as mentioned in Section~\ref{sec:problem_adblockers}.
Thanks to our approach, users can continue receiving profile-based targeted ads\footnote{As users are okay to be tracked on some categories of web pages. They still can receive useful ads targeted to those interest-categories of users.} 
on categories of web pages they accept to be tracked.

Our approach does not follow an ``all-or-nothing" policy as opposed to current ad blockers.
We allow users to select the categories of web pages that they consider sensitive and do not want to be tracked and receive ads on.
Additionally, users can be even more selective in the sense that they can block a web page without blocking the entire category.
For instance, if a particular web page hosts intrusive ads that belongs to a non-sensitive category, users can selectively penalize it while allowing ads on other web pages in the same category.
This feature may encourage publishers not to include intrusive ads on their web pages.

It is worth noting that our proposed approach is not a replacement but complements other initiatives of the ad industry~\cite{lean,acceptableads}.
While these initiatives aim to improve the quality of ads delivered to users, we give the onus to users to decide where they want to accept tracking and third-party ads. 
We emphasize that users who outrightly reject all ads are not considered.
Other economic models (for example, subscription-based access to the content) must be put in place for such users~\cite{Brave16Web}.

MyTrackingChoices is a tool that implements our approach and is described in more details in the following section.

\subsection{Implementation Details}
MyTrackingChoices is developed as Chrome extension and is available to download from Chrome Web Store~\cite{MtcChromeStoreUrl}.
In terms of functionalities, it allows users to block trackers (and thus, third-party ads) based on pre-defined categories of web pages or on a per web page basis.
If users do not agree with the categorization of a web page by MyTrackingChoices, they are allowed to change the category of that web page.
Users can also view the list of third-party domains present on a web page.

MyTrackingChoices extension on Chrome currently supports web pages in English, French, Spanish and Italian. 
This is due to the limitation of our current ``Categorizer" module (described next) but we plan to extend support for more languages in the near future.

MyTrackingChoices is composed of three main modules namely ``Categorizer", ``Policy module", and ``Blocking module".
Below is the detailed description of these modules. 
We note that, in the rest of the paper, the terms ``web page" and ``URL" are interchangeably used.

\subsubsection{Categorizer}
This module classifies the pages visited by users into a predefined set of topic of interests. 
The module employs a 2-level hierarchical taxonomy, composed of 32 \textit{top-level categories} and 330 \textit{bottom-level categories or subcategories}.
For the sake of usability, in the current version of the extension, we display and let users interact only with top-level categories.
However, depending on the feedback of users, e.g., if they find the top-level categorization too coarse, we can further allow users to interact with \textit{bottom-level categories}. 
Top-level categories can be found in the Table~\ref{tab:TopLevelCategories}, whereas subcategories corresponding to three of these categories are shown in Table~\ref{tab:SubLevelCategories} in the appendix of this paper.

The categorization algorithm integrated into our system is partly inspired by the methodology presented in~\cite{Kae11LSDM} for classifying non-textual ads into interest categories. 
The algorithm also builds on the taxonomy available with the Firefox Interest Dashboard plug-in~\cite{FID14plugin} developed by Mozilla.

Our categorizer relies on two sources of previously-classified data. 
First, a list of URLs, or more specifically, domains and hostnames,
which is consulted to determine the page's category. 
Here, it is worth noting that this list of URLs only contain domains that have all web pages belonging to a particular category.
For instance, techcrunch.com contains all articles related to ``technology \& computing" category.
Second, a list of unigrams and bigrams~\cite{Manning99B} that is used when the URL lookup fails. 
The former list is justified by the fact that a relatively small part of the whole Web accounts for the majority of the visits~\cite{amrit_gsb_privacy}. 
It is evident that pre-categorized lookup requires few computational resources on the user's browser and can be more precise. 
The latter list, on the other hand, is justified as a fall-back and allows us to apply common natural-language heuristics to the words available in the URL, title, keywords and content of the web page.

To build the first list of URLs, we incorporate Alexa.com's 500 top Web sites for almost each of the top-level categories.
The list also includes the domains and hostnames (around seven thousand) classified by Mozilla's Interest Dashboard plug-in~\cite{FID14plugin}.

On the other hand, in the list of unigrams and bigrams, we have approximately 76\,000 entries for English language. 
Three additional lists of unigrams and bigrams, although with a fewer number of entries, are also available for French, Spanish and Italian.
To compile all these words lists, we have built on the following data:

\begin{itemize}
  \item a refined version of the categorization data provided by the Firefox Interest Dashboard extension;
  \item a subset of the English terms available at WordNet 2.0~\cite{Miller95ACMCOM} for which the WordNet Domain Hierarchy~\cite{Magnini00LRE,Bentivogli04MLR} provides a domain label;
  \item a subset of the terms available at the WordNet 3.0 Multilingual Central Repository~\cite{Gonzalez12GW}, to allow the categorization of Web sites written in the aforementioned languages;
  \item and the synset-mapping data between the versions 2.0 and 3.0 of WordNet~\cite{Daude03RANLP}.
\end{itemize}

The categorizer module resorts to word lists only when the hostname and domain are not found in the URL list.
When this happens, the algorithm endeavors to classify the page by using the unigrams and bigrams extracted from the following data fields: URL, title, keywords and content of the page. 
Depending on the data field in question, the categorizer assigns different weights to the corresponding unigrams and bigrams. 
In doing so, we can reflect the fact that those terms appearing in the URL, the title, and especially the keywords specified by the publisher (if available), are usually more descriptive and explanatory than those included in the body of the page.

As frequently done in information retrieval and text mining, our web page classifier also relies on the Term Frequency-Inverse Document Frequency (TF-IDF) model~\cite{Salton75COM}. 
Said otherwise, we weigh the resulting category(ies) based on the frequency of occurrence of the corresponding unigrams and bigrams in the web page, and on a measure of their frequency within the whole Web.

Based on the categorization using words, we have a score for each category depending on the algorithm above described.
To select the categories that best represent a web page, we select all the categories that have a score larger than a threshold value, where threshold ($T$) is defined as
$$
T = \alpha \times \textnormal{(maximum score - average score),}
$$
where  $\alpha$ is a constant and its value is chosen based on how much weight we want to give to accuracy versus completeness (false positives versus true negatives).
In our current implementation, we choose $\alpha$ equal to 0.3.
If there are less than three such categories, the web page is classified into all these categories.
However, if there are more than three such categories, we pick only three top categories and categorize the web page into them.
Thus, the maximum number of categories a page may belong to is upper bounded by 3.

For the sake of computational efficiency, the algorithm caches the categories derived from the user's last 500 visited pages. 
This way, when the user re-visits one of those pages, the interest-categories of these pages are obtained directly without needing to go through the whole process of categorization described above.

In terms of storage, the whole list of unigrams, bigrams and their corresponding IDF values occupies approximately 1 megabyte in compressed format. 
We believe this is an acceptable overhead in terms of the extension download size.

Lastly, a manual inspection of the categorization results for a small collection of web pages indicates that the algorithm works reasonably well. 
Experimental results will show later in Section~\ref{subsec:performance} that users are satisfied with our categorizer (only 18 web pages with erroneous categorization are reported from 12 users).
However, further investigation is required to evaluate the performance of the categorizer in a more rigorous manner.

\begin{figure}[t]
\centering
\fbox{\includegraphics[scale=0.28]{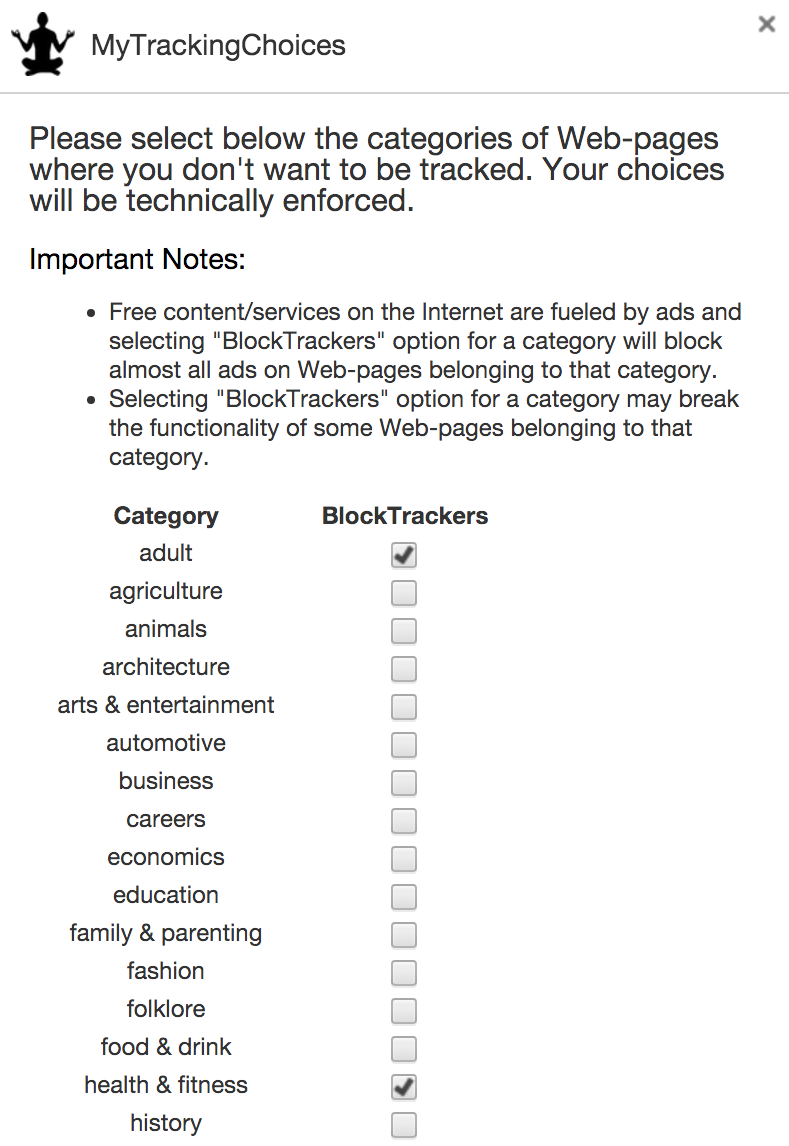}}
\caption{The configuration panel shown in this figure allows users to define per category blocking policies.}
\label{fig:CatPolicyPanel}
\end{figure}

\textbf{A potential approach to improve categorization.}
As seen in Section~\ref{sec:problem_adblockers}, a web page should be categorized based on its content and not on its domain name except if all the pages in the domain belong to a particular category.
This is because a particular domain can host web pages from several different categories.
For example, domains belonging to the ``news" category, such as cnn.com, usually include web pages from a variety of categories (sports, economy, politics, health, travel, religion, etc.).
To this end, our tool takes a best-effort approach to categorization by deriving semantic information from several data associated to the web page such as URL, title, keywords and content.
Unfortunately, this approach does not guarantee accuracy.
A simple yet effective solution to this problem would be to recommend web developers to include the category of the web page in the \texttt{<head>...</head>} tag of the HTML code.
One potential way to encourage Web developers in doing so would be to include our recommendation in the W3C best practices.
Apart from the obvious advantage of a more accurate categorization, this solution also eliminates the extra overhead (in terms of computation, memory and network bandwidth) incurred by the tool to categorize a web page.
However, enforcing developers to include (correct) categories in web pages is a challenging task.

\begin{figure}[t]
\centering
\fbox{\includegraphics[scale=0.35]{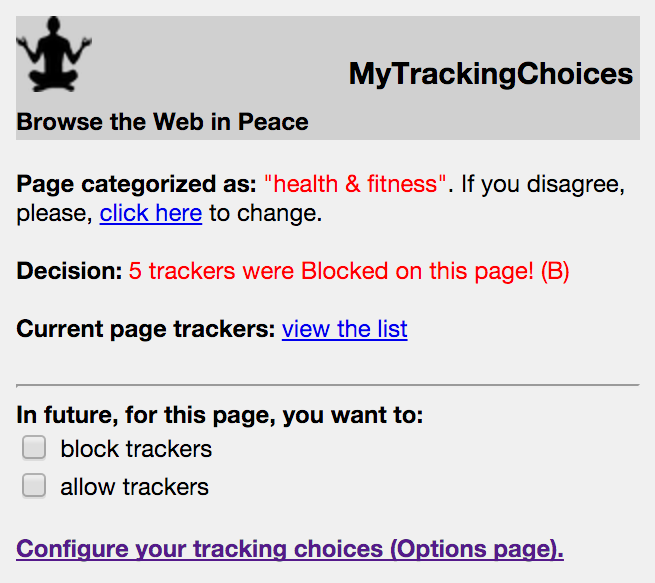}}
\caption{Screenshot of the popup page. It shows the category(ies) of the current visited page along with the decision about blocking or allowing the trackers.
Depending on the first experience with the web page, users are also given the option to block or allow trackers the next time they visit that web page.}
\label{fig:screenshot}
\end{figure}

\subsubsection{Policy module}
This module is responsible for applying the blocking policies defined by users.
Policies can be defined either on per category basis or on per web page basis.

To simplify the implementation, our extension allows users to define only per category blocking policies.
That is, instead of specifying categories of web pages where tracking should be allowed and categories of web pages where tracking should be blocked, we just enable the latter blocking declaration.

Fig.~\ref{fig:CatPolicyPanel} shows the configuration panel where users may configure the per category blocking.
After having installed the extension and before the extension starts functioning, users are presented with the options panel so that they can configure their choices (by default, no categories are blocked). 
Configuration of blocked categories can also be performed at a later stage by clicking on ``configure your tracking choices" option in the popup page of MyTrackingChoices.

Contrary to the per category policies, users are allowed to configure both block and allow per URL policies.
This is to allow users to be more granular if they are occasionally not satisfied with per category blocking policies.
For example, in a scenario where a user has blocked a category of web pages but wants to support a specific web page in that category by receiving ads.
Fig.~\ref{fig:screenshot} shows how the popup page of MyTrackingChoices looks like.
It provides users with options to define per URL policies.

The operation of this module is described next.
When a user visits a page, the module waits for the categorizer to send the topic category of the web page browsed.
Being equipped with the URL as well as the category, the module decides whether the network connections of third-party domains should be blocked or not.
We emphasize here that in case of a conflict between per web page and per category based policy, per URL based policy prevails.
In case a web page is categorized in more than one category, if any of the category is selected by the user to be blocked, third-party network connections on that web page are blocked.

\subsubsection{Blocking module}
This module is responsible for blocking network connections of third-party domains if a user has selected to block them.
There are two main tasks.
First, finding third-party domains and then, second, to block them.
To find third-party domains, we just need to check if the domains of network connections match with the domain the user typed in the address bar.
If it is not the same, then, such domains can be considered as third-party domains.

The next task is to block network connections to such domains.
However, blocking network connections of all third-party domains may break the functionality of some web pages.
This is because of the fact that some web pages download useful content from other domains (a different domain belonging to the first-party domain or a content provider or some other domain).
To counter this problem, existing ad blockers keep the list of domains that deliver ads or are tracking users.
More specifically, this list of domains is also accompanied with some regular expressions that are used to pinpoint only some network connections from those domains.

MyTrackingChoices takes a different approach to find which network connections of third-parties to block so that the functionality of the web page is not broken.
Instead of collecting all tracking and advertising domains (which are huge and dynamic), we build a list of domains that are essential for the functionality of a web page.
In our experience, such a list is smaller and easier to maintain than the list of tracking and advertising domains.
Therefore, we only keep the list of such domains~\cite{list_allowed_domains}.
We block all other domains that do not belong to this list and are classified as a ``tracker" based on a heuristic described below.

We classify a third-party domain as a ``tracker" if it is present on three or more different domains that a user visited in the past. 
This implies that the extension becomes fully functional only after a user visits a couple of web pages after installation.
This heuristic is employed so that we do not need to maintain a list of third-party domains that are specific to a first-party domain to deliver content.
For example, ``lemonde.fr" uses ``lemde.fr" domain to deliver content to their web pages. 
However, this domain is only specific to ``lemonde.fr" and is probably used by Le Monde only for this purpose.
Therefore, such domains are not classified as ``trackers".
As we do not need to keep the list of such domain specific third-party domains, this helps us to keep our list of useful domains reasonably small.


\section{Economic Impact}
\label{sec:economic_impact}
This section assesses if our approach can contribute to decreasing the impact of ad blocking on the Web economy.
This assessment is based on the data collected from users of MyTrackingChoices.
In the following, we first describe our dataset and the ethical considerations regarding the data collection and use.
And in the sequel, we measure the economic impact of MyTrackingChoices.

\textbf{Dataset.}
Before the extension is put on the Chrome Web Store\footnote{\url{https://chrome.google.com/webstore/detail/mytrackingchoices/fmonkjimgifgcgeocdhhgbfoncmjclka?hl=fr}}, it was distributed among friends and colleagues for beta testing.
Data from these users are not collected and used in our study.
The collected data is from those users of MyTrackingChoices who downloaded it directly from Chrome Web Store. 
Such users either found the extension on the Chrome Web Store independently or came to know about it through various publicilty channels (word-of-mouth, twitter, facebook, etc.).
The dataset, used for different analyses in this paper, includes data from January 11 to February 20, 2016.

The dataset contains 96 users\footnote{We 
consider one installation as one user. 
However, it is technically possible for a user to uninstall the extension and install it again. We assume that users did not do this.}.
These 96 users are those who browsed 20 or more web pages and configured at least once their options regarding blocking of third-party domains (either on a per category or on a per web page basis).
Users, who did not configure blocking feature at all, have not been considered in this study because the installation of the extension serves no purpose in this case.
The choice of minimum browsing of 20 web pages is somewhat arbitrary but the intuition behind this is not to take into account the preferences of users who did not considerably use the extension.
Such users may further distort the ensuing analysis.

Based on data uploads from users to our server, we find that different users used MyTrackingChoices extension for different time intervals ranging from a minimum of around 6 hours to a 
maximum of 44 days.
Even though it is possible for users to continue using the extension and just opt-out of data upload, we consider that they stopped using the extension when no data is received from users.
There were 7.29\% users (7 users out of 96 users) who used our extension for less than 1 day.
A vast majority of users used it for more than 1 day.

\textbf{Ethical considerations regarding data collection.}
The collected data is considered to be pseudo-anonymized because it does not contain any information that can be directly used to re-identify users.
It contains hashes and categories of web pages visited by the users along with third-party domains and URLs of iframes present on those web pages.
As a part of responsible disclosure, users were informed that the extension is developed by Inria as a research project and collected data will be used to study the impact of MyTrackingChoices on the Web economy. 
We provided users with the option to opt-out of data collection as well as delete previously uploaded data to our servers.
Furthermore, we assured users not to make an attempt to re-identify them and also, not to share the data with any outside party.
MyTrackingChoices privacy policy\footnote{Available at \url{https://myrealonlinechoices.inrialpes.fr/privacy_policy.html}} clearly specifies all the details regarding our data collection, storage,
processing and sharing.

\subsection{Analysis}

The analysis is done under the reasonable assumption that users know how to properly use the extension.
To ensure proper functioning of the extension, we had put a clear notice\footnote{Here 
is the excerpt from our web page: ``Please uninstall other Adblockers or disable blocking functionality in other anti-tracking extensions like Ghostery, DisConnect, PrivacyBadger etc. 
before installing MyTrackingChoices. Otherwise, your choices would not be completely respected."} on our project web page\footnote{\url{https://myrealonlinechoices.inrialpes.fr/}} as well as extension download page~\cite{MtcChromeStoreUrl} 
on Google Chrome Store
to uninstall other ad blockers or disable blocking functionality in other anti-trackers.
Indeed, such extensions could intervene in the functionality of MyTrackingChoices and therefore, user choices may not be completely respected.

In the following, we present results cumulated over all the users of \emph{MyTrackingChoices} since our goal is to measure the global effect on the Web economy.
As we are interested in different aspects related to different categories of web pages, most of our results do not focus on users but categories of web pages.

\begin{figure}[t]
\centering
\includegraphics[width=\columnwidth]{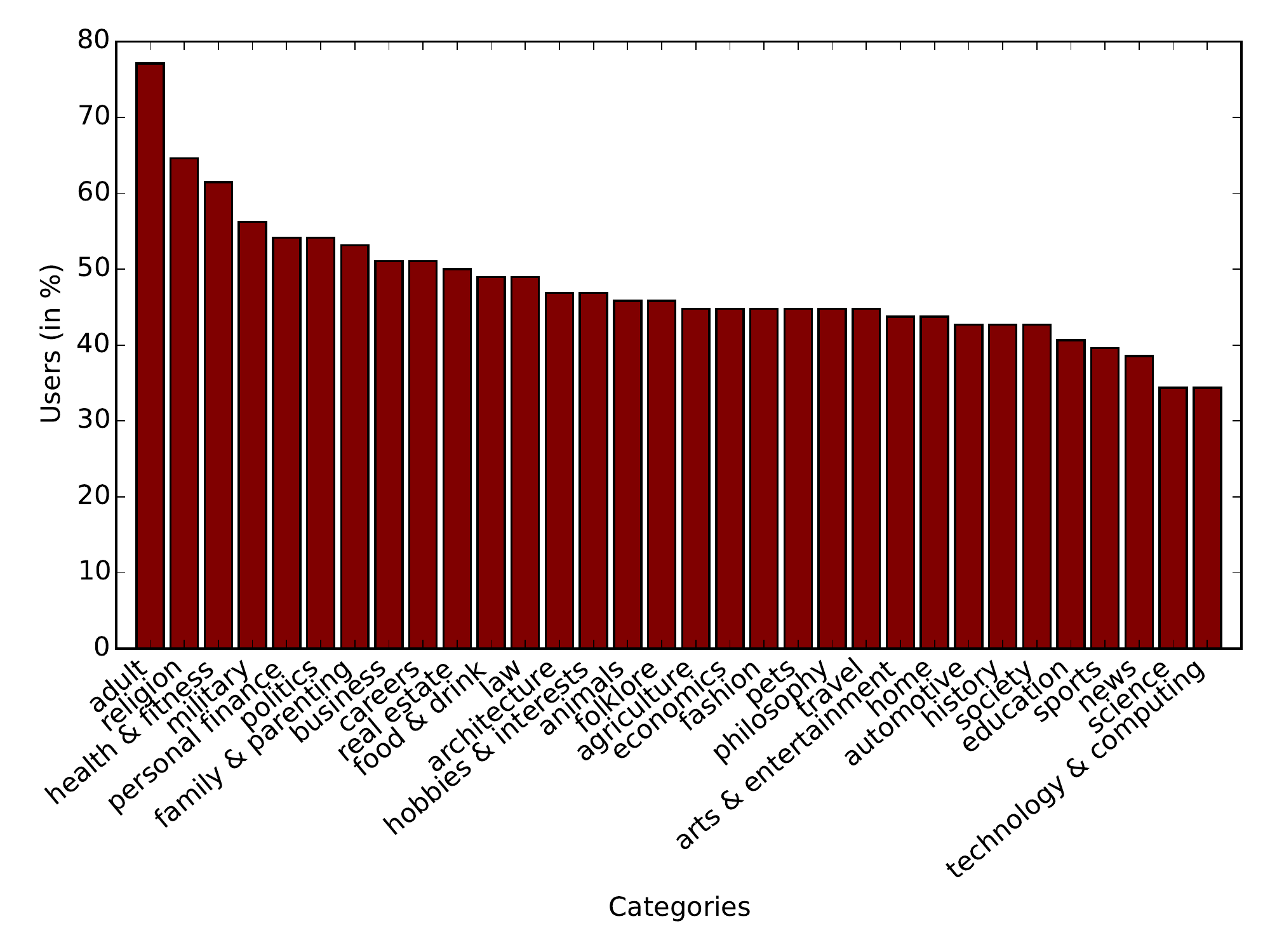}
\caption{Percentage of total users blocking a category. Total number of users considered in this study are 96.}
\label{fig:blocked_cats_sorted}
\end{figure}

To measure the impact on economy, we first start by looking at number of users who actually exercised fine-granular choices.
We find that 69.98\% users accept the presence of third-party domains and thus, tracking and hosting ads in some categories of web pages that they do not consider much sensitive.
On an average, a user blocked 15.28 categories (almost half) out of a total of 32 categories.
The median of the number of blocked categories is 11.
These bare statistics suggest that the impact on the Web economy due to ad blocking can be reduced if users are provided with such fine-grained choices.
However, we need to keep in mind that such a solution is for users who block ads because of privacy reasons, who are not against ads and do not want to be tracked on some categories of web pages that they consider sensitive.

We also note that 30.02\% (29 users out of a total of 96) users selected all categories to block network connections of third-party domains.
These are probably the users that are either against tracking altogether or probably the ones who do not want to see ads at all.
In fact, blocking trackers on all visited web pages will block almost all ads as most ads are delivered today through third-party domains present on a web page.
As these users simply do not want to exercise fine-granular choices, providing them such an option will not be help survive the ad-based Web economy.

Next, we are interested in knowing categories of web pages that are most or least blocked by users.
This should enable us to know the categories that are the most sensitive to users (where most users do not want to be tracked) and categories that users care the least with respect to tracking.
Equipped with this information, in future, we can propose to users the categories of web pages to block by default and let user tweak them if they are not satisfied.

Fig.~\ref{fig:blocked_cats_sorted} presents the categories blocked by users in the order of the most blocked to the least blocked.
We note here that the three most blocked categories are ``adult", ``religion" and ``health \& fitness".
These categories are, in fact, also considered as \emph{sensitive} by European Data Protection Law~\cite{handbook_data_protection}.
The three least blocked categories are ``technology \& computing", ``science" and ``news".
This result confirms that users are more concerned about sensitive categories~\cite{handbook_data_protection}.
Therefore, it actually makes sense to provide them with such an option, i.e., to block trackers based on the categories of web pages.

As we now know the most and the least blocked categories by users, it is interesting to study the browsing habits of users (category of web pages they browse the most/least) 
to get more insights about the impact on the web economy.
The total number of browsed web pages by these 96 users are 86,922, out of which 12,543 are distinct.
\begin{figure}[t]
\centering
\includegraphics[width=\columnwidth]{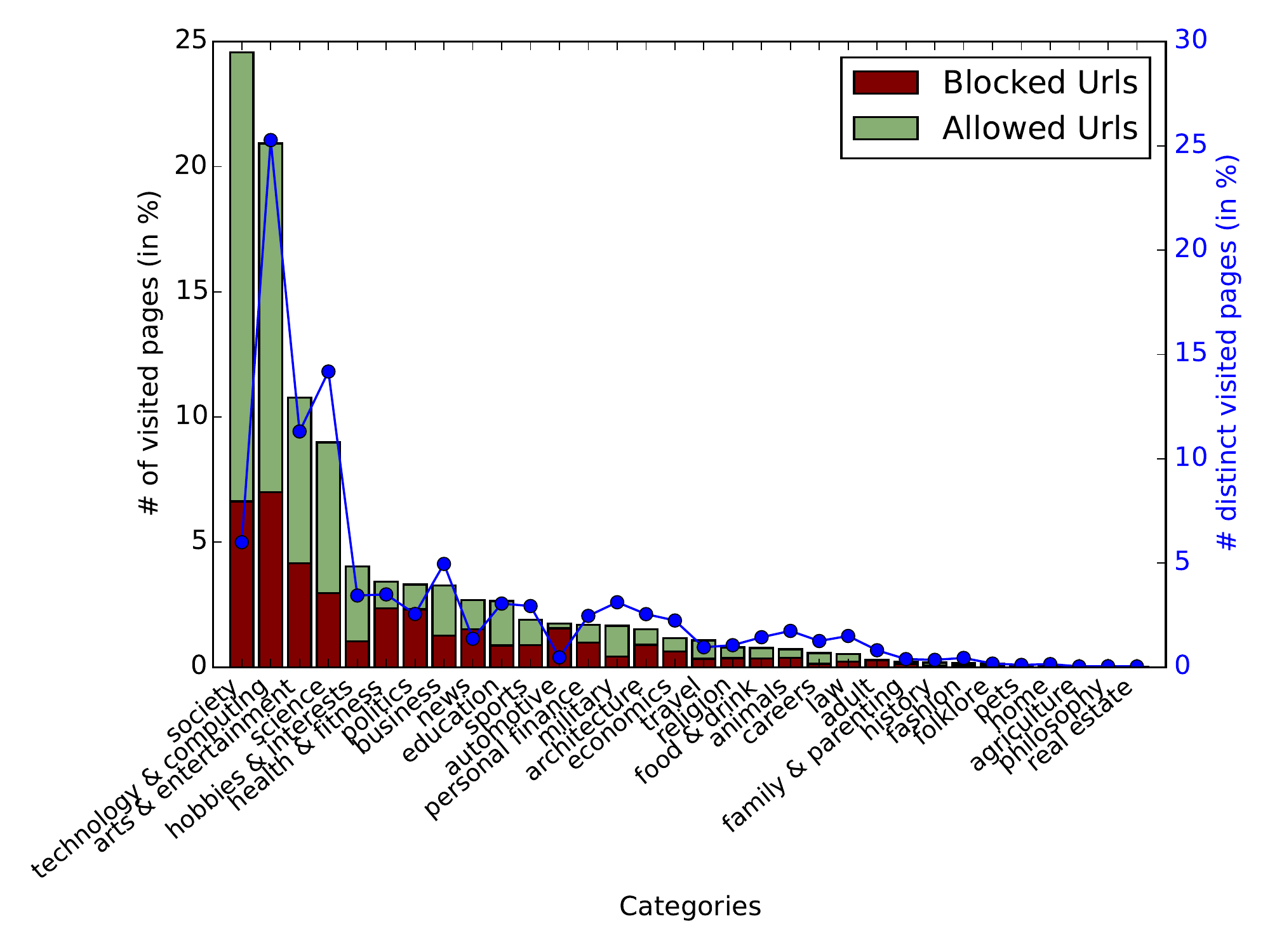}
\caption{Distribution of blocked and allowed visited pages by category. Right side y-axis presents distribution of distinct visited web pages by category. 
The total number of browsed web pages are 86,922 whereas the total number of distinct web pages browsed are 12,543.}
\label{fig:dist_visited_pages_by_cat}
\end{figure}
Fig.~\ref{fig:dist_visited_pages_by_cat} presents the distribution of total and distinct visited pages per category.
This also shows the distribution of blocked versus allowed pages in each category.
In order to detect any bias in the distribution of browsed web pages (owing to small number of users or type of users), we check it with other studies and the pattern seems to resemble well~\cite{goel2012does, Kumar2010}.

We find that the top-5 most browsed categories by our extension users are ``society", ``technology \& computing", ``arts \& entertainment", ``science" and ``hobbies \& interests".
Here it is worth noting that web pages browsed in these 5 categories constitute for roughly 65\% of the total browsing activity.
However, none of these top-5 most browsed categories (Fig.~\ref{fig:dist_visited_pages_by_cat}) are in the top-5 of the most blocked (Fig.~\ref{fig:blocked_cats_sorted}).
This explains why there are more allowed than blocked web pages in the top-5 categories of Fig.~\ref{fig:dist_visited_pages_by_cat}.

There are more blocked than allowed web pages,  in the Fig.~\ref{fig:dist_visited_pages_by_cat}, for the categories that are the most blocked in Fig.~\ref{fig:blocked_cats_sorted}.
As the most blocked categories are least browsed and the most browsed categories are the least blocked, the effect on the economy is less damaging.
In total, only 33.19\% of browsed web pages were blocked.
This shows that the number of blocked web pages are far too less than the ones where trackers (and thus, third-party ads) are allowed.

Until now, we studied the effect on the economy due to fine-grained blocking options on per web page basis.
This already indicates that MyTrackingChoices can reduce the impact of ad blockers on the Web economy.
However, as we also collect the source URLs of all iframes included in the web pages browsed by users, this allows us to measure how many iframe-based ads actually got blocked.
As iframes are generally used to deliver ads from third-party domains directly into a web page, we can make an accurate estimate of the impact of \emph{MyTrackingChoices} on the Web economy.
It is worth noting that this study excludes non-iframe based ads as well as those delivered directly by the publisher.
However, as per our experience, percentage of such ads is very low. 

In total, we find 958 different domains delivering iframes over all 96 users.
However, since all iframes do not necessarily contain ads, we first filter those iframes that are used to deliver ads.
This filtering is done based on the list of advertising domains from Mozilla Focus project and their partner Disconnect~\cite{disconnect_ad_domains}.
After this filtering, we find a total of 115 different ad domains and a total of 27,861 iframes or ads served by these domains.
Table~\ref{tab:top_ad_domains}, in the appendix of this paper, contains top 40 ad domains that delivered ads through iframes.

\begin{figure}[t]
\centering
\includegraphics[width=\columnwidth]{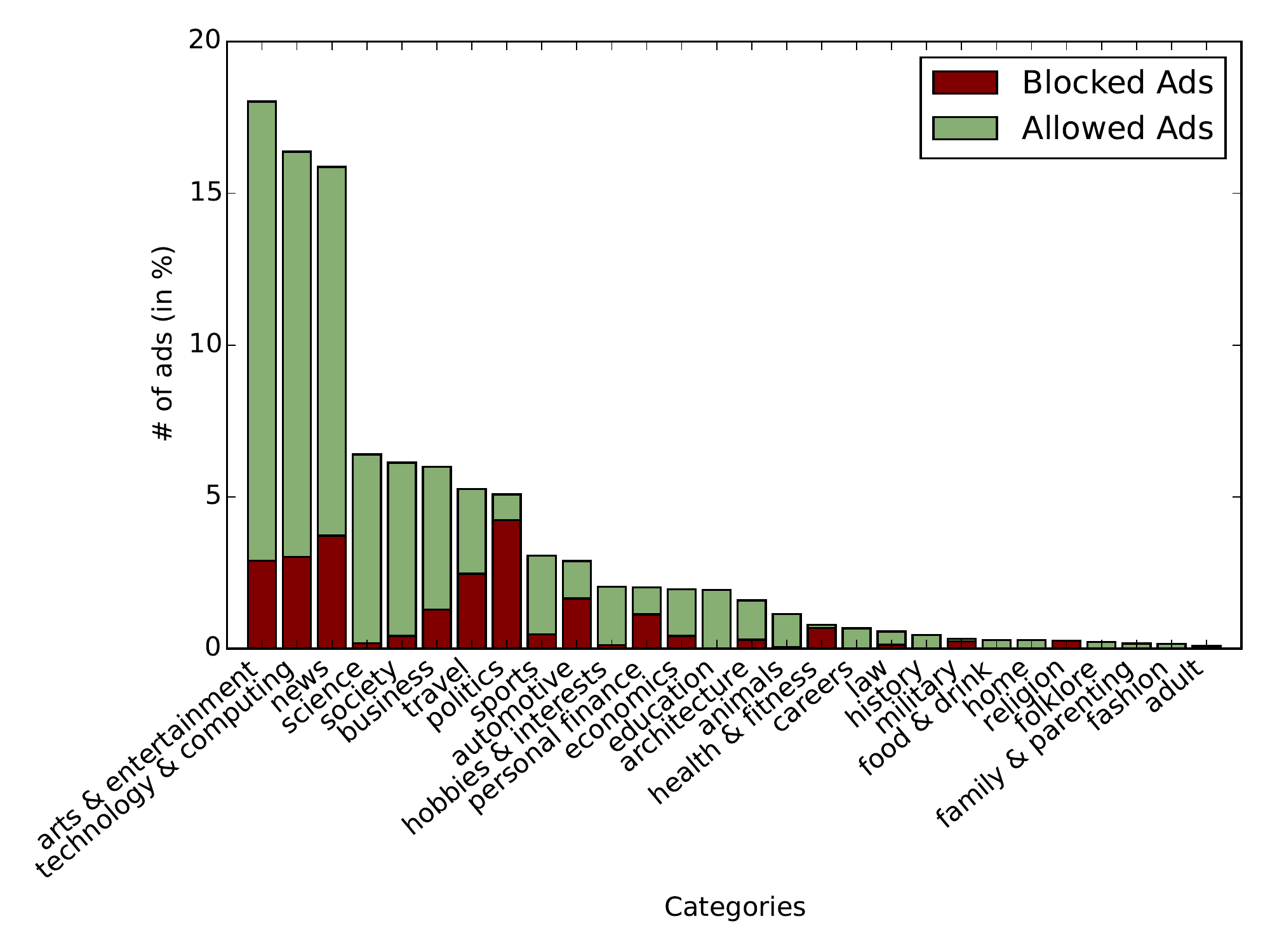}
\caption{Distribution of ads by category and on the fact if they are blocked or allowed. Total number of ads are 27,861.}
\label{fig:dist_ads_by_cats}
\end{figure}
Fig.~\ref{fig:dist_ads_by_cats} presents per category distribution of ads.
For each category, we also have the fraction of ads that are allowed or blocked due to blocking policies specified by users.
The categories of web pages which host most of the ads are ``arts \& entertainment", ``technology \& computing" and ``news".
As expected, we find that the percentage of ads the most blocked are in categories that are most blocked by users, for example, ``adult", ``health \& fitness", ``religion", ``politics", ``personal finance".
However, as web pages belonging to these categories are comparatively less visited by users as well as such web pages have comparatively lesser number of ads, the total number of ads blocked by all users of \emph{MyTrackingChoices} is small.
In total, only 23.8\% of all ads are blocked.

\textbf{Conclusion.}
Assuming that our collected data is representative of users blocking ads due to privacy reasons, 
we believe that the economic impact of ad blocking exerted by privacy-aware users can be significantly reduced if we provide users with such a fine-grained control.
As the loss due to ad blocking was 21.8 billion dollars in the year 2015 according to PageFair 2015 report~\cite{pagefaire_report_2015}, this could have been reduced to 14.7 billion dollars with the use of \emph{MyTrackingChoices}.
We estimate this reduction in the economic loss based on the facts that only 23.8\% ads are blocked by \emph{MyTrackingChoices} users and 50\% of users block ads due to privacy concerns~\cite{pagefaire_report_2015}.

\section{Usability and Performance Evaluation}
\label{sec:usability_performance}
This section assesses the usability and performance of MyTrackingChoices.
We report upon the pseudo-anonymous data collected from users of our extension.

\subsection{Usability}
\label{subsec:usability}
As usability is an essential factor for a wide scale adoption of a tool, we evaluate two of its important aspects.
Before going into further details, we reiterate that current anti-trackers allow 
users to block ads based on a per URL basis whereas MyTrackingChoices lets users specify
their blocking policies based on categories of web pages.
Bearing in mind this distinction, our first evaluation criterion considers the 
frequency with which users need to change the two blocking policies.
Under this criterion, we compare the blocking policies used by current anti-trackers and MyTrackingChoices.
It is evident that category-based blocking should outperform URL-based blocking.
However, for the sake of completeness and accuracy, we measure the performance of MyTrackingChoices in this regard.

We note that in certain circumstances, category-based blocking can be too coarse.
For example, a user may not want to block a specific web page in a blocked category or vice versa.
To this end, MyTrackingChoices, offers an additional possibility to be even more granular and block or allow on per web page basis.
Our second criterion captures the usage of this feature.
This criterion is evaluated by counting the number of users who exercise such a granular feature and at how many web pages this feature is used.

In the following, we measure the afore-described criteria on our dataset.

First, we study how users' tracking choices evolve over time.
Specifically, we want to know if users are satisfied by just configuring once their tracking choices when they install the extension or 
they change their tracking choices post-installation.
\begin{figure}[t]
\centering

\begin{tikzpicture}
\pie{81/Once , 14/Twice , 3/Thrice , 2/More}
\end{tikzpicture}

\caption{Percentage of users modifying the ``category-wise blocking of trackers setting" different number of times.}
\label{fig:evolution_blocked_cats}
\end{figure}
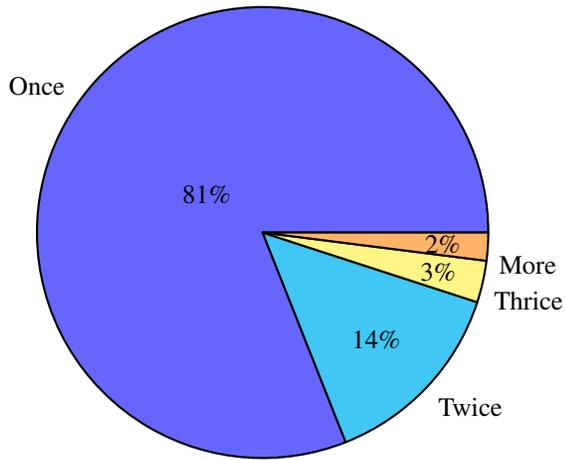
Fig.~\ref{fig:evolution_blocked_cats} presents the percentage of users who change their tracking choices.
We note here that 81\% users configure their choices once and then, do not feel the need to change it.
This signifies that blocking based on categories is stable over time and this makes it highly usable.
Additionally, we highlight that almost all (16 out of 18) remaining users added one or more categories, 
while only 2 users removed an already blocked category.

Next, we measure how users exercised per URL blocking option.
We find that 52 users out of a total of 96 users, i.e., 54.17\% users, used this functionality.
This clearly suggests that such an option is useful to have.
In total, web pages where URL-based blocking feature is used, constitute for 1.01\% (127 web pages out of 12,543) of total web pages browsed by these 52 users.
This is not a considerable percentage but we need to keep in mind that all such users also exercised their choice regarding blocking trackers based on the categories of web pages.
They usually need to exercise per web page blocking in case they are not satisfied with their per category blocking decision.

\begin{figure}[t]
\centering
\includegraphics[width=\columnwidth]{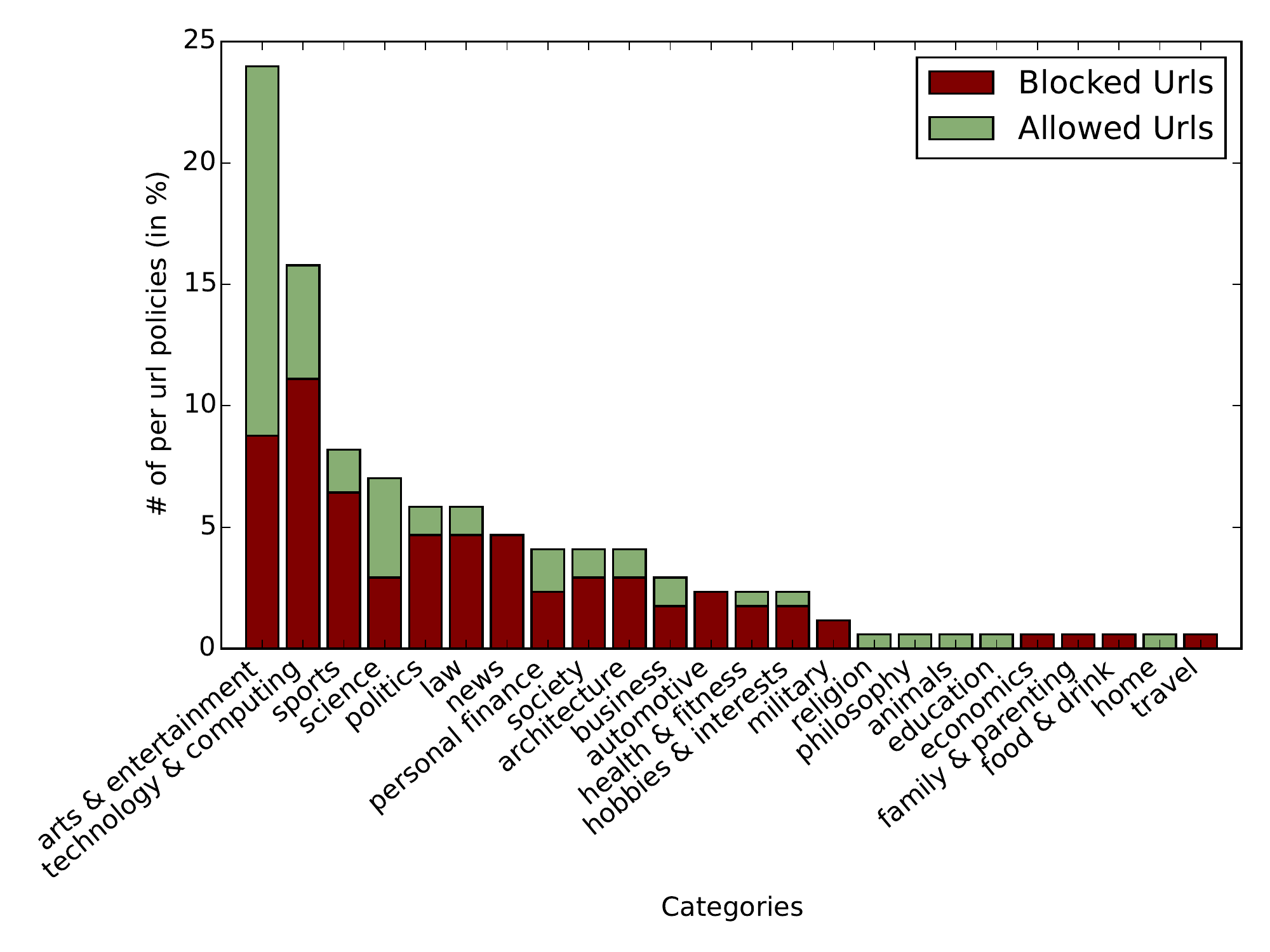}
\caption{Distribution of per URL defined policies by category.}
\label{fig:dist_per_url_policy_by_cat}
\end{figure}
Fig.~\ref{fig:dist_per_url_policy_by_cat} presents the distribution of per URL granular policies exercised by these 52 users over different categories of web pages.
It is worth noting that the top-5 categories for which per URL policies are defined, none of them belongs to the top-5 categories blocked by users (Fig.~\ref{fig:blocked_cats_sorted}).
This means that the web pages which contribute the most to the per URL policies are those that belong to non-sensitive categories.
Also, we note that 37.42\% of per URL policies are defined to allow trackers on different web pages where tracking is blocked as per their category.
This leads us to the conclusion that some users are willing to accept ads on web pages belonging to non-sensitive category and indeed want to support publishers.
However, at the same time, we also notice that trackers/ads are blocked on a per URL basis for web pages that are not blocked based on their categories. 
This implies that users do block trackers and thus, eventually ads on certain web pages even if they did not consider their categories sensitive.
This user behavior suggests that these web pages may contain intrusive ads and therefore, users opted for blocking of third-party domains.


\subsection{Performance}
\label{subsec:performance}
We measure the performance of \emph{MyTrackingChoices} in various respects. 
First, we measure the performance of our extension in terms of correctness of categorization algorithm and proper functioning of the web pages.
We find that incorrect categorization is reported for only 18 web pages (out of a total of 12,543 distinct web pages browsed by all the 96 users) by 12 users.
This shows that users are mostly satisfied with the categorization of web pages by our extension.
Also, seven users reported 23 web pages whose functionality broke due to MyTrackingChoices.
Most of the times, it was because of blocking of content-provider domains and we corrected the problem by updating the list of allowed domains.

\begin{figure*}
\begin{subfigure}[b]{0.33\textwidth}
  \includegraphics[width=\textwidth]{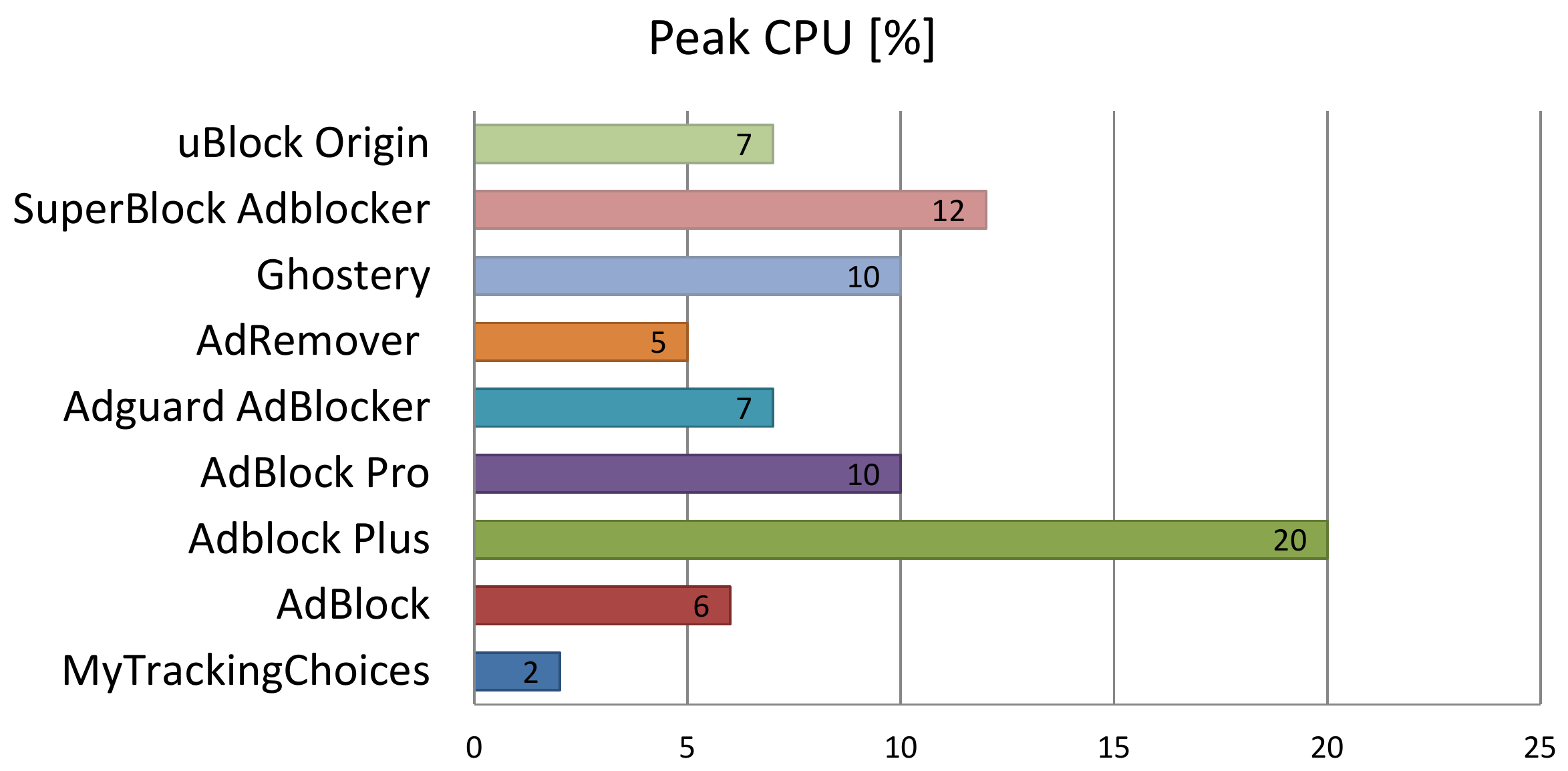}
\end{subfigure}%
\begin{subfigure}[b]{0.33\textwidth}
  \includegraphics[width=\textwidth]{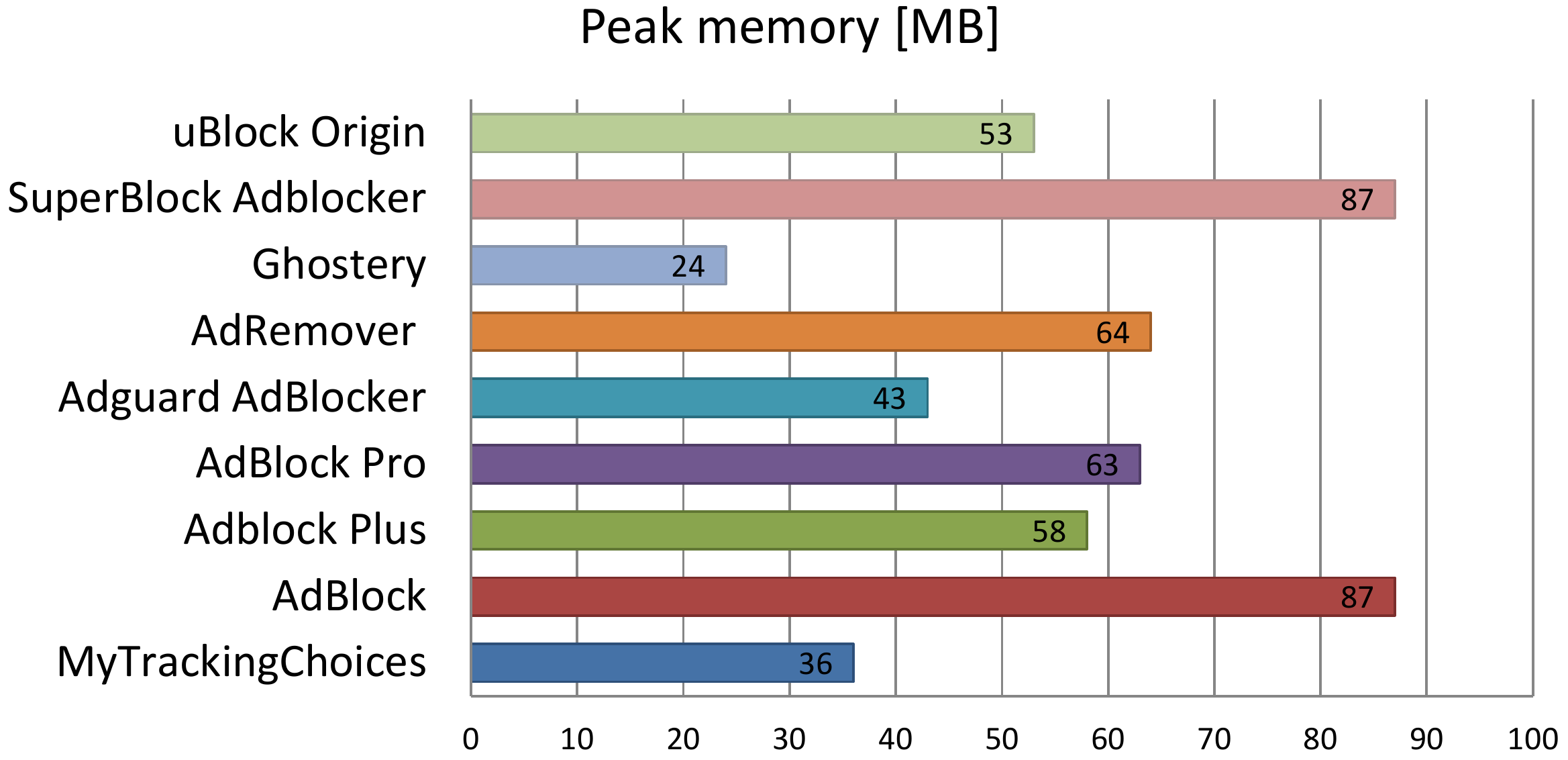}
\end{subfigure}
\begin{subfigure}[b]{0.33\textwidth}
  \includegraphics[width=\textwidth]{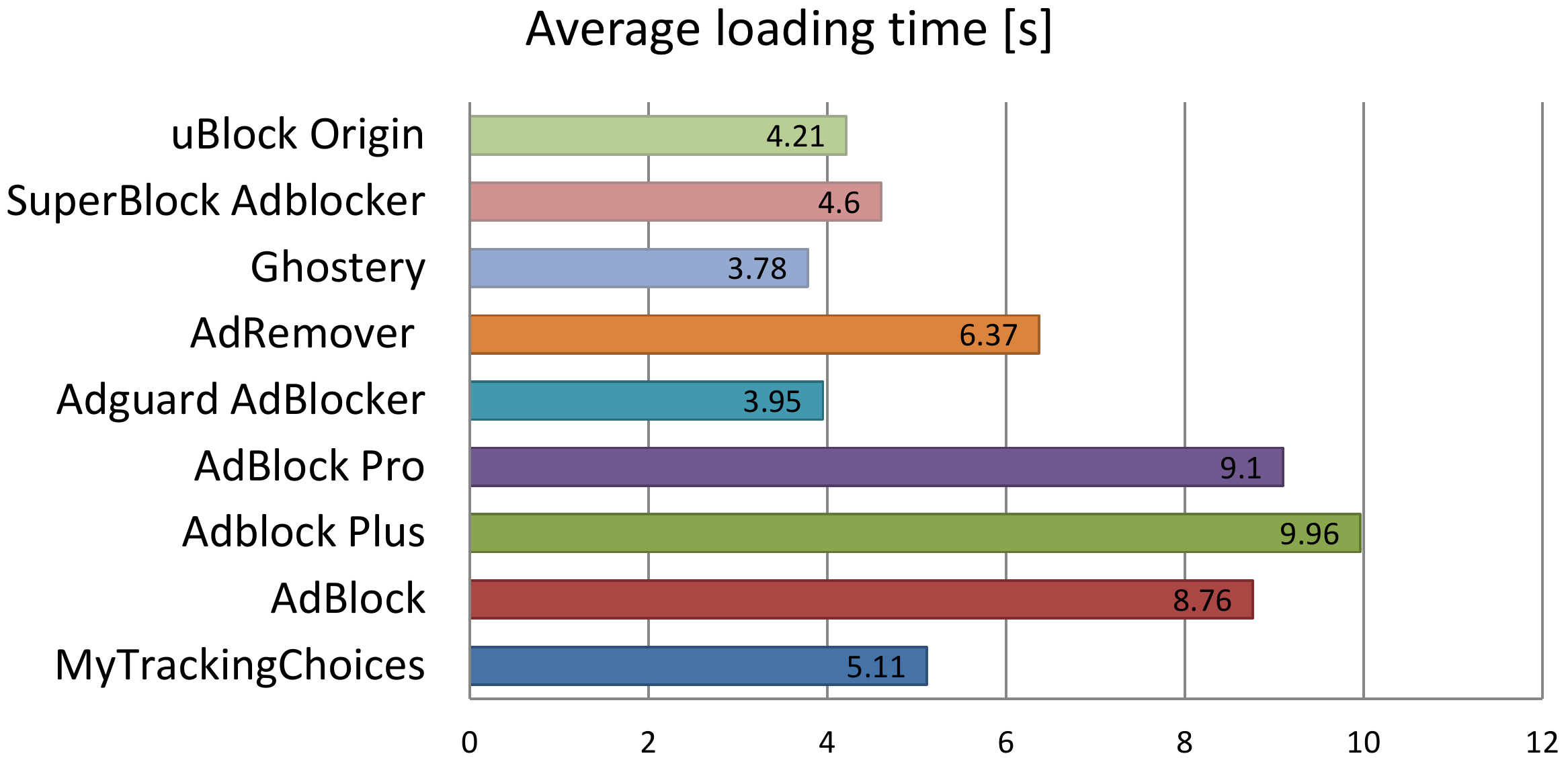}
\end{subfigure}

\begin{subfigure}[b]{0.33\textwidth}
  \includegraphics[width=\textwidth]{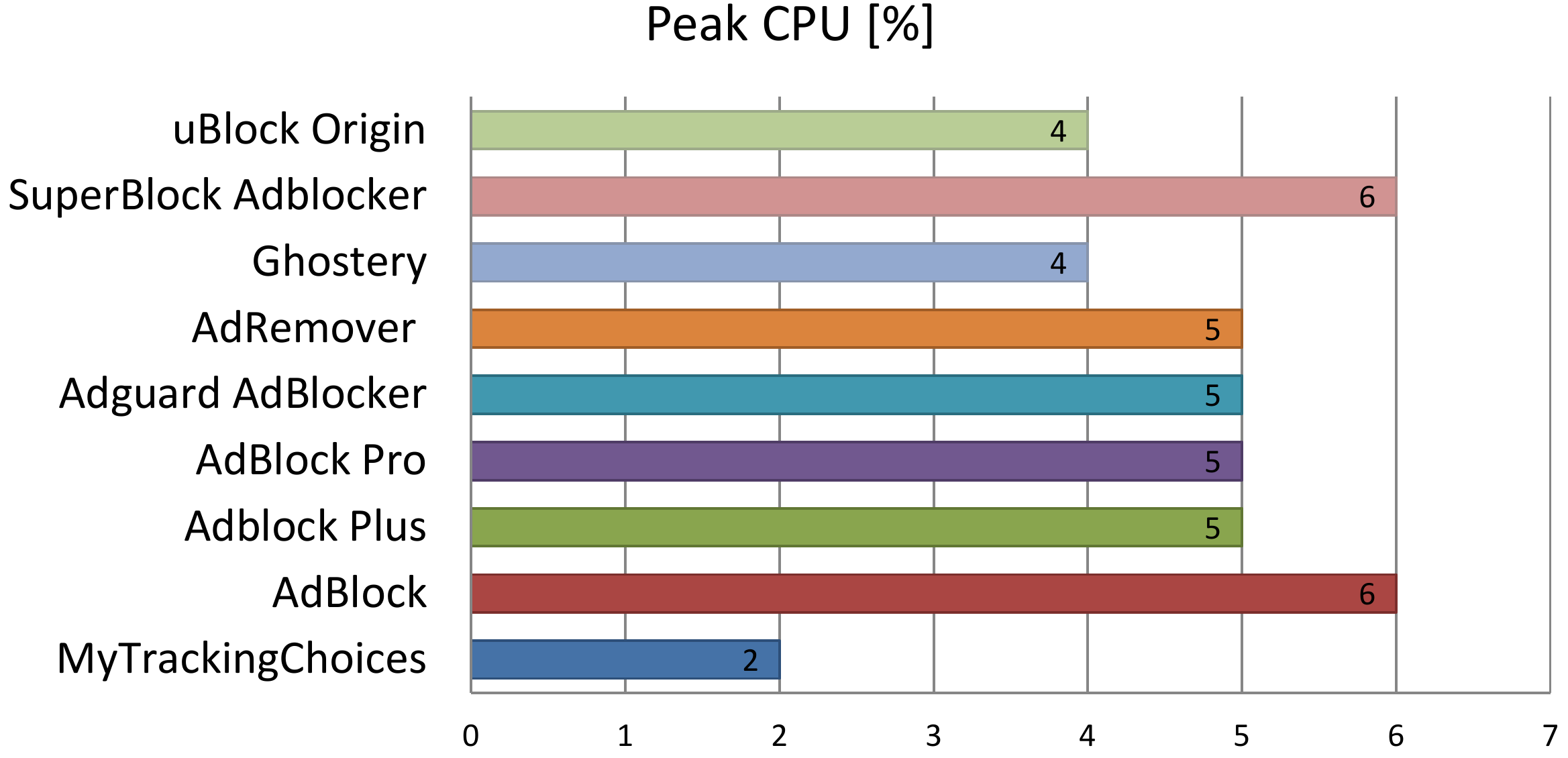}
\end{subfigure}%
\begin{subfigure}[b]{0.33\textwidth}
  \includegraphics[width=\textwidth]{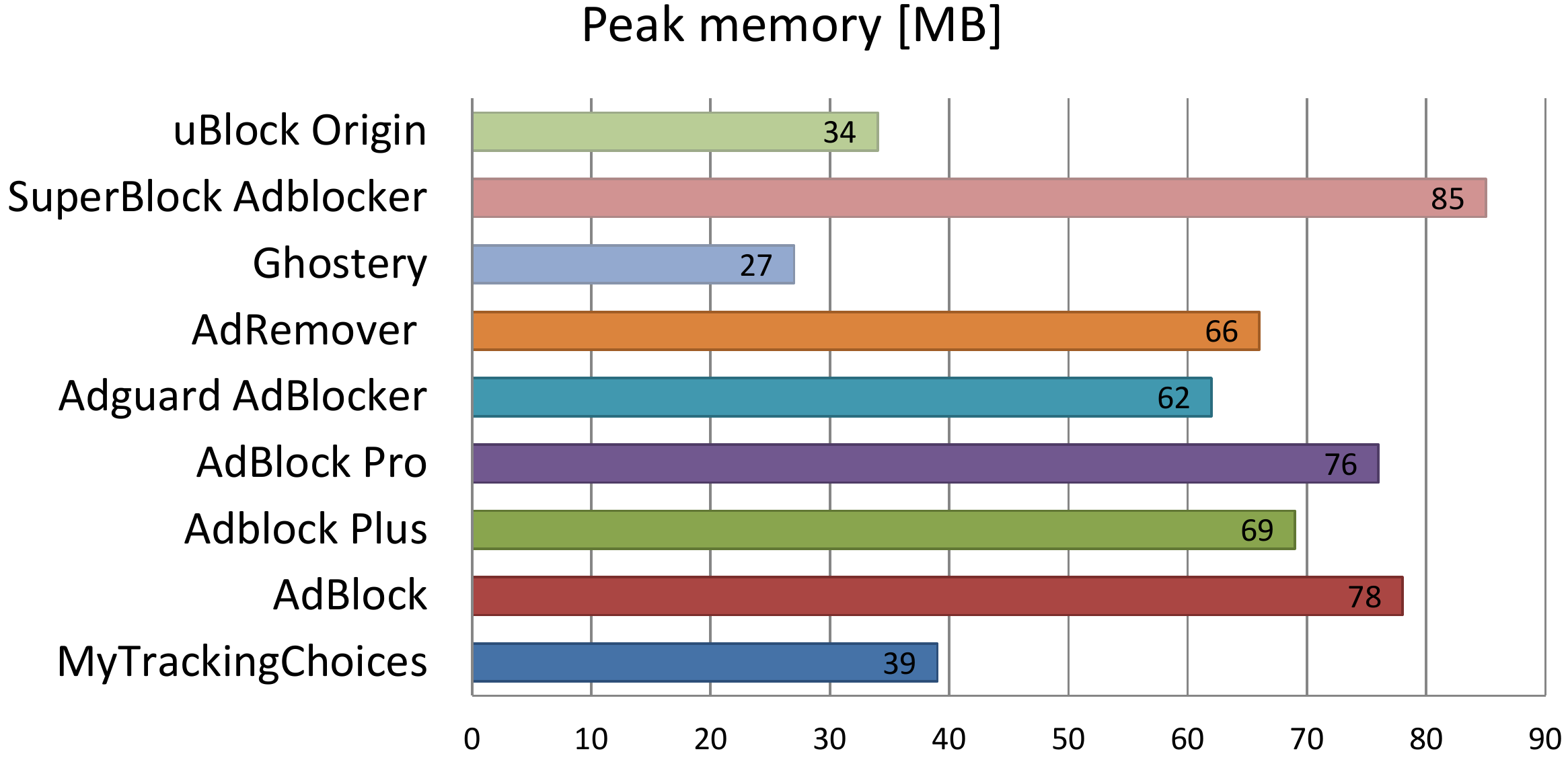}
\end{subfigure}
\begin{subfigure}[b]{0.33\textwidth}
  \includegraphics[width=\textwidth]{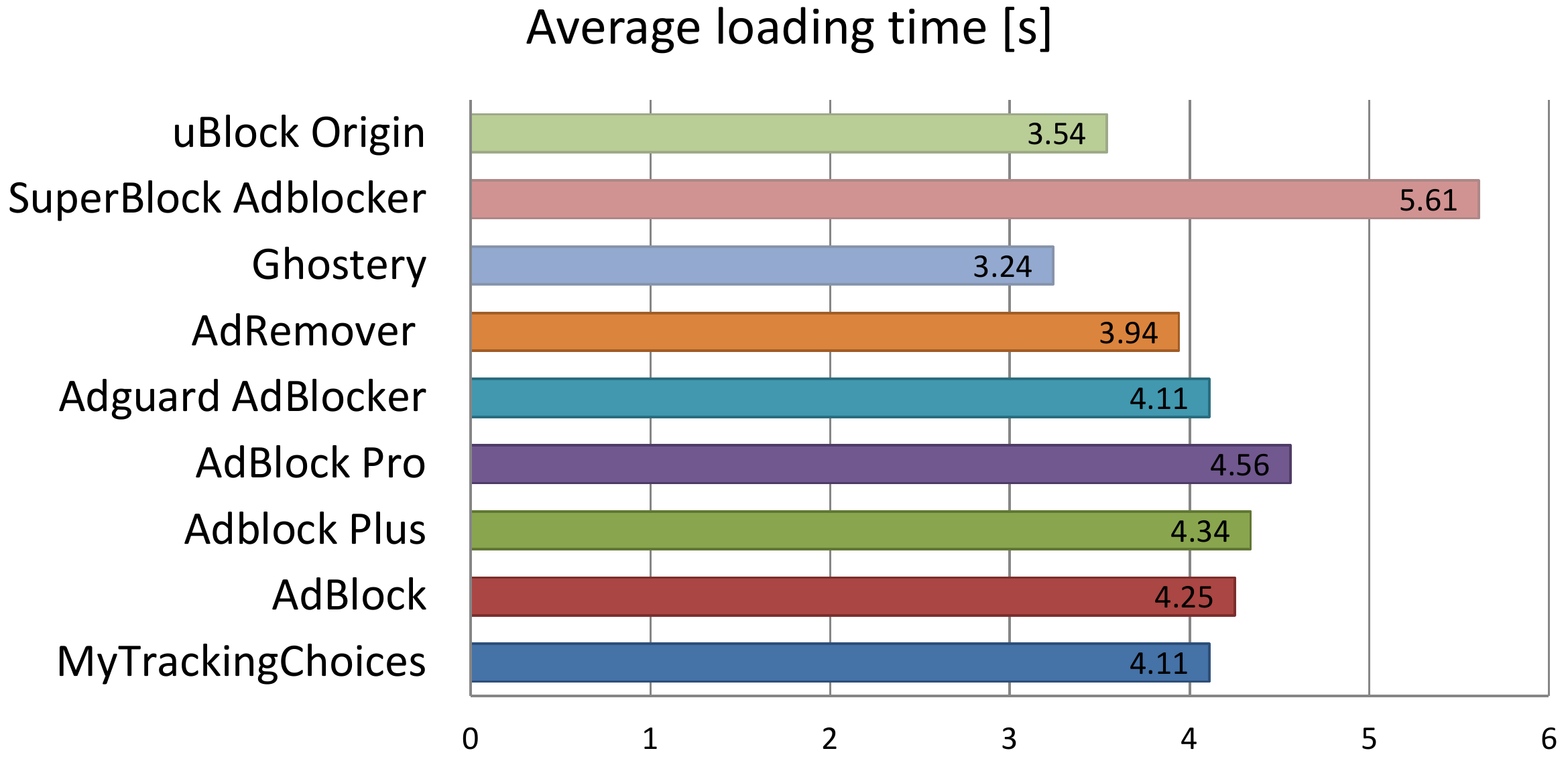}
\end{subfigure}

\begin{subfigure}[b]{0.33\textwidth}
  \includegraphics[width=\textwidth]{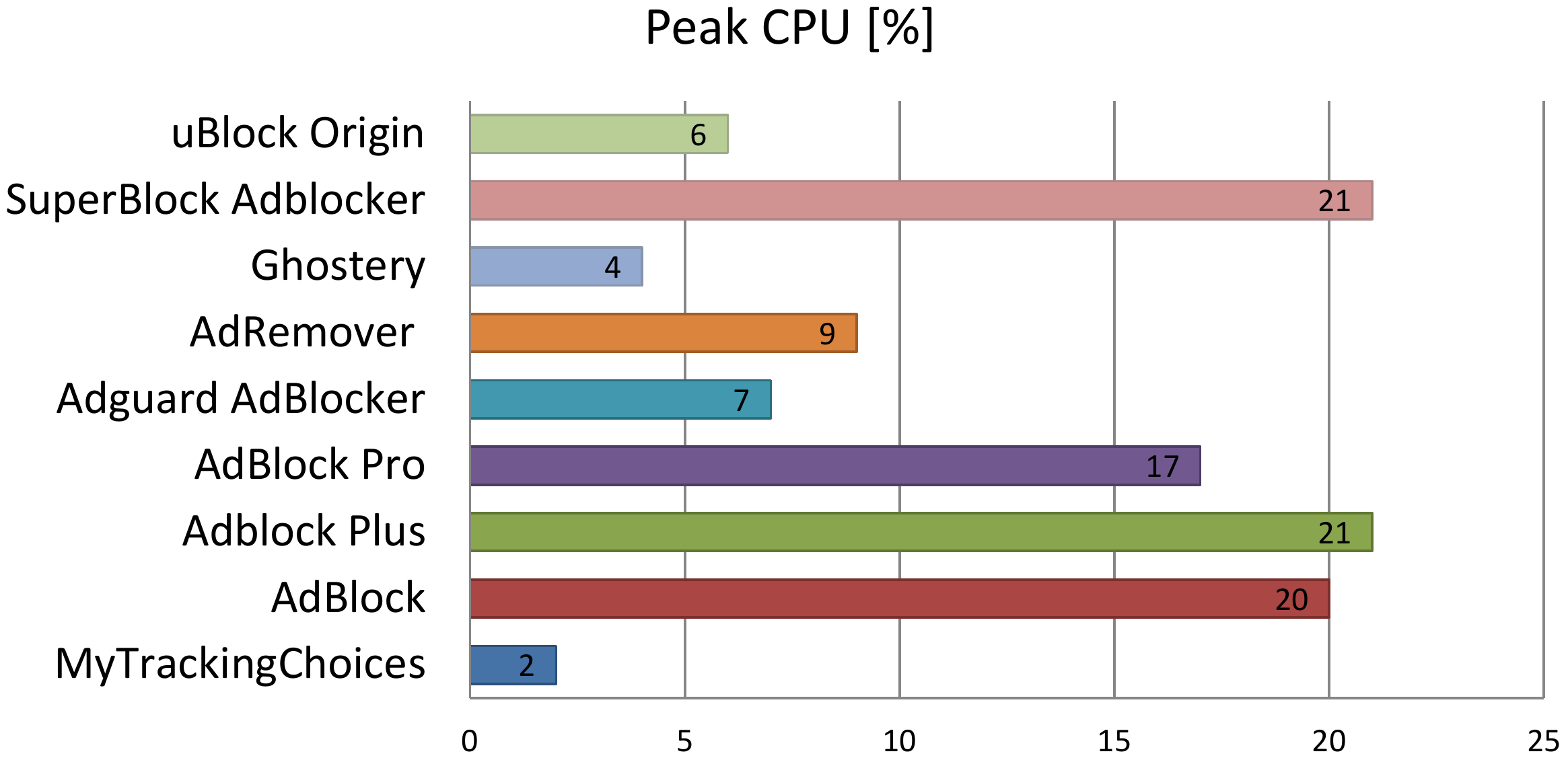}
\end{subfigure}%
\begin{subfigure}[b]{0.33\textwidth}
  \includegraphics[width=\textwidth]{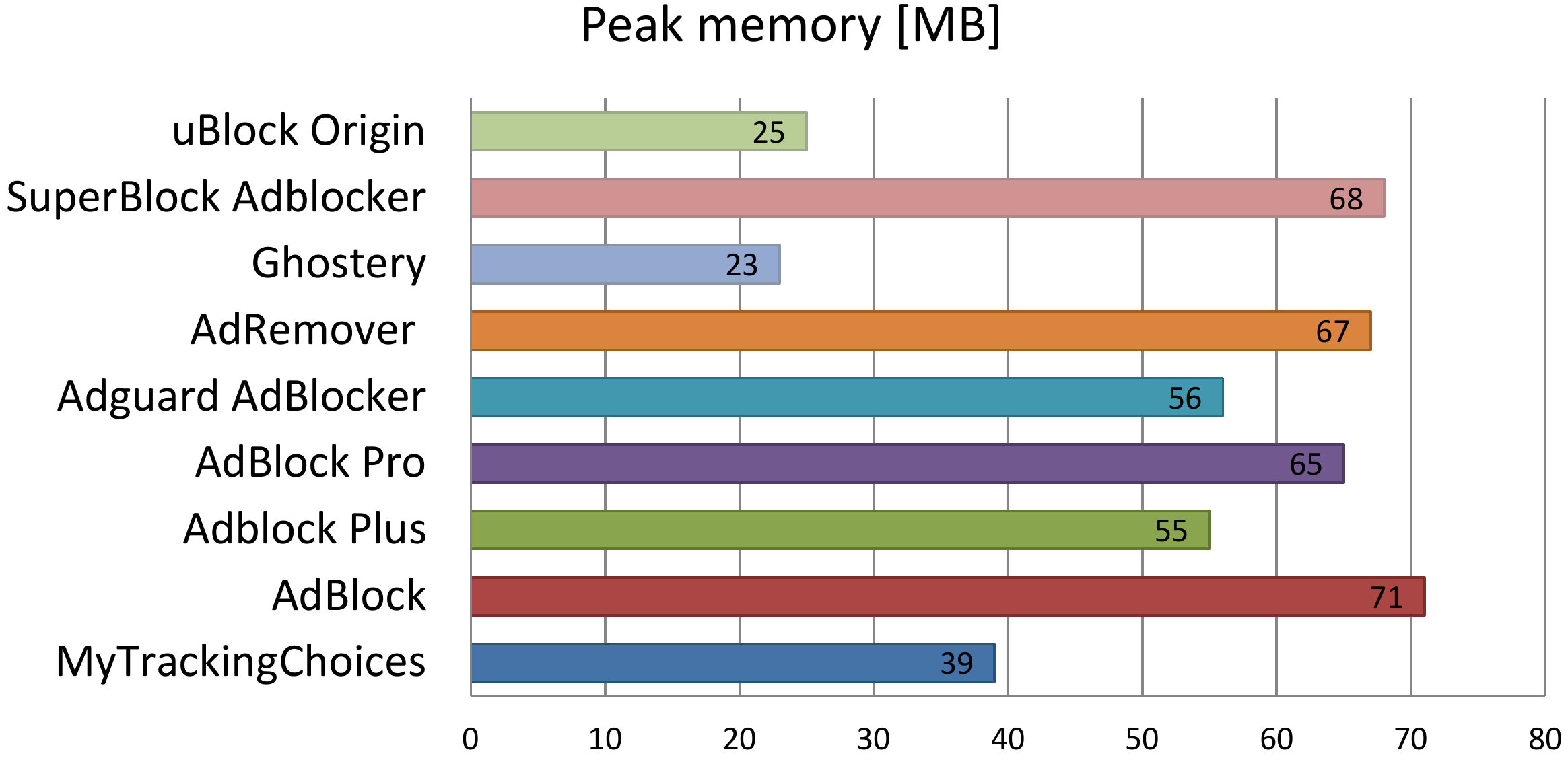}
\end{subfigure}
\begin{subfigure}[b]{0.33\textwidth}
  \includegraphics[width=\textwidth]{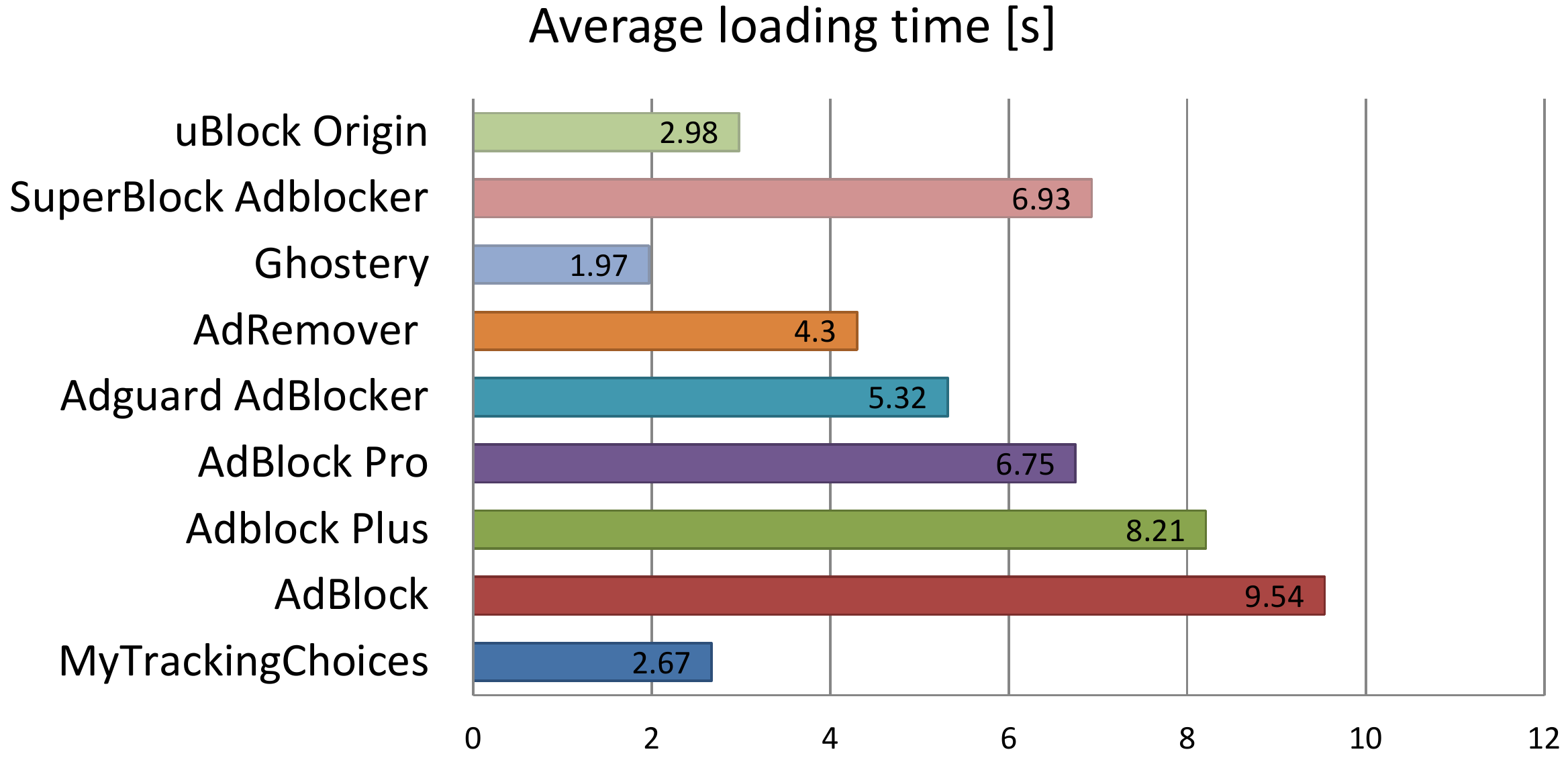}
\end{subfigure}

\caption{The first row shows the peak CPU usage (in \%), peak memory usage (in MegaBytes) and average loading time (in seconds) for home page of CNN website (cnn.com). 
The second and third rows show the same information for home pages of Le Monde (lemonde.fr) and New York Times (nytimes.com) respectively.}
\label{fig:performance_benchmark}
\end{figure*}

Next, we assess our extension in terms of average page-loading time, 
maximum memory usage and maximum CPU usage, and compare these results with some of the most prominent ad-blockers and anti-trackers. 
This is necessary because our extension provides certain privacy guarantees in terms of user tracking but it comes at the cost of processing and computational overhead.

Fig.~\ref{fig:performance_benchmark} presents the benchmark analysis of this comparison.
The results of this benchmark analysis have been obtained after ten consecutive visits to these three popular web pages: 
cnn.com, lemonde.fr and nytimes.com.
The main conclusion that can be drawn from this figure is that our extension, despite its higher complexity, outperformed most of the tools evaluated in this study. 
In terms of page loading time, MyTrackingChoices is close to Ghostery and uBlock Origin, the two tools which led this aspect. 
As for memory usage, our extension required between 37 and 39 MB, ranking third after Ghostery and uBlock Origin in lemonde.fr and nytimes.com, and second after Ghostery in cnn.com. 
Finally, with just 2\% of processing usage, our extension outperformed all tested tools in terms of CPU consumption. 
In a nutshell, the performance of MyTrackingChoices was above the average for those three web pages, 
and this was despite the inclusion of more sophisticated and advanced privacy features.

\section{Presence of Trackers}
\label{sec:measurement_trackers}

In this section, we study the prevalence of trackers on the Web.
We first discuss how this study is conducted and then, present the measurement results.

If a third-party domain is present on a web page, we assume that it is tracking users.
Moreover, a third-party domain is classified as ``tracker" only if it is found to be tracking users on three or more web pages (as described in Section~\ref{sec:problem_adblockers}).
The consequence is that first two presence of a third-party domain does not suffice for the domain to be classified as a ``tracker".
In addition, we exclude from our study web pages that are blocked by users.
This is because if the initial request to a tracker is blocked, none of the consequent requests, that would have been made otherwise, can be captured.

Users in our dataset were tracked 239,420 times by a total of 588 different trackers.
The average number of trackers, over all users, per web page is 3.2 with a standard deviation of 5.1.
However, the maximum number of trackers present on a web page can be as high as 98 in some cases whereas the minimum is zero.

\begin{figure}
\centering
\includegraphics[width=\columnwidth]{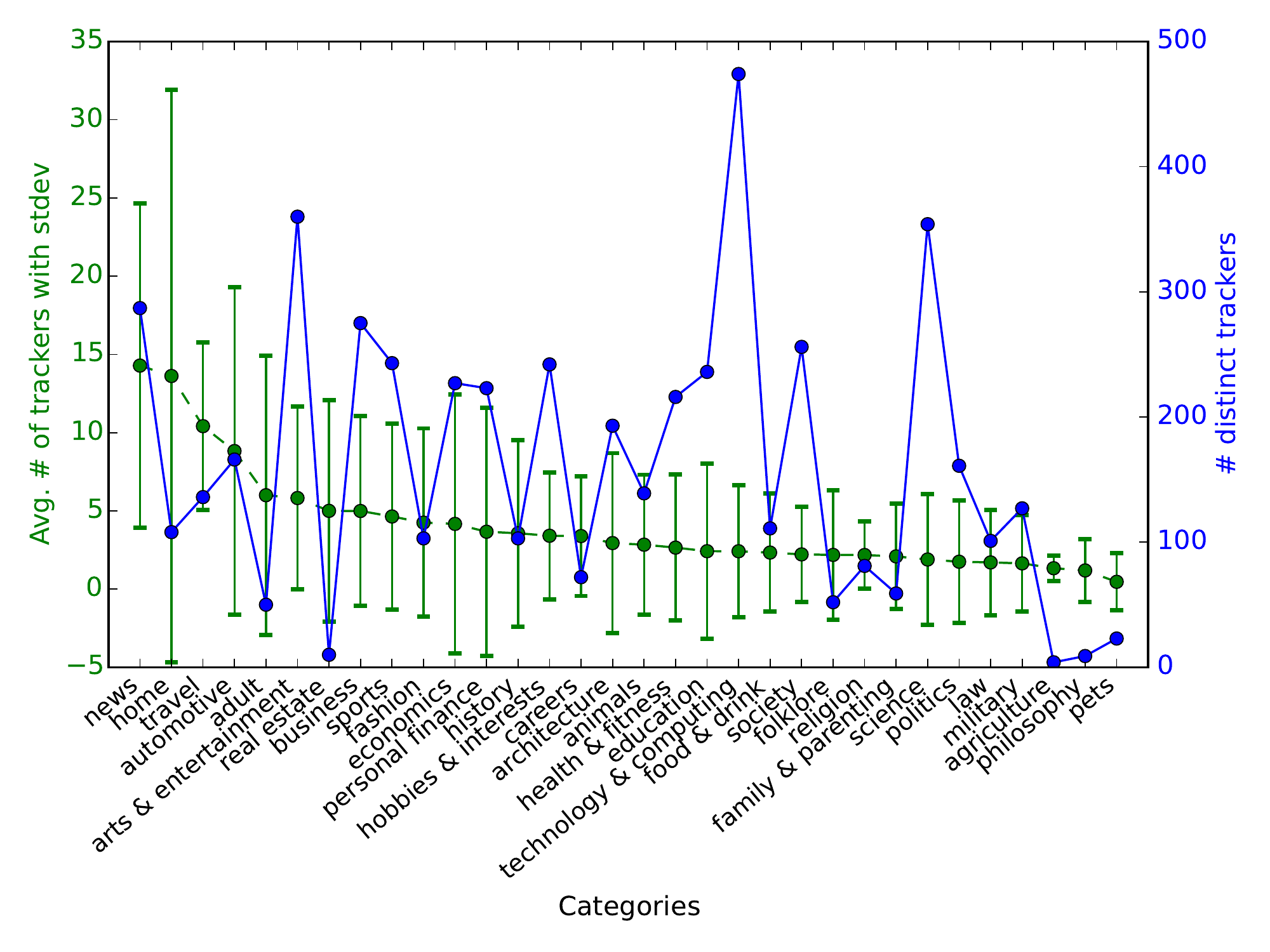}
\caption{Distribution of trackers by category. Left-side y-axis (in green) represents average number of trackers with standard deviation whereas
right-side y-axis represents number of distinct trackers.}
\label{fig:dist_trackers_by_cat}
\end{figure}
Fig.~\ref{fig:dist_trackers_by_cat} presents the distribution of trackers over different categories.
From this figure, we note that web pages in the ``news" category contain the maximum number of trackers on average.
However, if we contrast it with number of ads received on web pages in the ``news" category (Cf. Fig.~\ref{fig:dist_ads_by_cats}), 
we find that the web pages in the ``news" category did not receive the maximum number of ads.
This suggests that web pages in the ``news" category include many additional trackers other than those related to ads.
In fact, the average number of trackers in the ``news" category is 14.28 with a standard deviation of 10.37.
This is much higher than the global average of 3.2 trackers per page.

From the same figure, it is interesting to note that web pages in the ``technology \& computing" 
have the most number of different trackers, close to 500, even though the average number of trackers per web page is quite small. 
In fact, web pages categorized as ``news" are tracked by 284 different trackers even though the average number of trackers per page is the highest.
This result suggests that the trackers are most spread in web pages belonging to ``technology \& computing" category.

We refer interested readers to Table~\ref{tab:frequent_trackers} in the appendix of the paper which shows the list 
of frequent trackers overall and in some of the sensitive category web pages.

\section{Conclusion and Future Work}
\label{sec:conclusion}

Considering the poor ad-experience on the part of users, ad blockers and anti-trackers have become an indispensable tool.
They provide absolute privacy to users at the cost of an adverse impact on the Web economy.
In this paper, we proposed a new approach that provides a balance between user privacy and the Web economy.
The proposed approach is user-centric and allows users to exercise fine-grained options.
It allows users to choose the categories of web pages that are privacy-sensitive to them and block trackers on such web pages only.
As a result, it helps users to control their profile collected by different third-parties, e.g., advertising companies, data brokers, insurance companies, etc.
 We believe that our proposed approach can provide the desired level of privacy while simultaneously supporting the ad-based economic model of the Web.
Overall, it can also help improving the user ad-experience and the quality of ads.

As a future work, we aim to extend our tool by proposing users an option to block annoying ads irrespective of where they appear.
Another line of work consists in letting user specify the maximum number of ads they are willing to receive on a web page belonging to a non-sensitive category web page.
Finally, we plan to block trackers by default on sensitive category web pages but users would have the option to tweak the default settings if they do not agree.

\section*{Acknowledgment}
\noindent
The writing quality of this paper is benefited by comments from Amrit Kumar.
This work is partially funded by the Inria Project Lab CAPPRIS.
J. Parra-Arnau is the recipient of a Juan de la Cierva postdoctoral fellowship, FJCI-2014-19703, from the Spanish Ministry of Economy and Competitiveness.

\bibliographystyle{unsrt}
\bibliography{paper}

\appendix

\begin{table*}[tb!]
\caption{Subcategories corresponding to three top-level categories.}
\label{tab:SubLevelCategories}
\centering
\footnotesize
\begin{tabularx}{.8\textwidth}{ >{\bfseries }cX}
\toprule
\textbf{Top-level category} & \makebox[9.6cm][c]{\textbf{Bottom-level category}} \\ [0.8ex]
\toprule
arts \& entertainment  & animation, celebrities, comics, design, fine art, humor, literature, movies, music, opera, poetry, radio, television,  theatre and video games. \\\midrule
health \& fitness & alternative medicine, anatomy, asthma, autism, bowel incontinence, brain tumor, cancer, cardiac arrest, chronic pain, cold \& flu, deafness, dental care, dermatology, diabetes, dieting, epilepsy, exercise, eye care, first aid, heart disease, HIV/AIDS, medicine, men's health, mental depression, nutrition, orthopedics, pediatrics, physical therapy, psychology \& psychiatry, senior health, sexuality, sleeping disorders, smoking cessation, stress, substance abuse, thyroid disease, vitamins, weight loss and women's health. \\\midrule
personal finance & banking, credit, debt \& loans, cryptocurrencies, financial news, financial planning, insurance, investing, retirement planning, stocks and tax planning.\\
\bottomrule
\end{tabularx}
\end{table*}

\begin{table*}[tb!]
\centering
\caption{Top 40 ad domains that delivered ads through iframes}
\label{tab:top_ad_domains}
\footnotesize
  \begin{tabular}{ccccc}
    \toprule
        
        doubleclick.net 35.62\% & googlesyndication.com 28.81\% & weborama.fr 3.80\% & 2mdn.net 3.66\% & rubiconproject.com 2.63\% \\
        criteo.com 2.22\% & outbrain.com 2.22\% & gemius.pl 2.14\% & betrad.com 1.78\% & pubmatic.com 1.46\% \\
        openx.net 1.35\% & criteo.net 1.33\% & serving-sys.com 0.92\% & casalemedia.com 0.79\% & bluekai.com 0.70\% \\
        amazon-adsystem.com 0.69\% & effectivemeasure.net 0.66\% & zedo.com 0.54\% & adnxs.com 0.53\% & tribalfusion.com 0.41\% \\
        demdex.net 0.38\% & mathtag.com 0.36\% & lijit.com 0.35\% & adverticum.net 0.34\% & infolinks.com 0.34\% \\ 
        doubleverify.com 0.33\% & quantserve.com 0.33\% & advertising.com 0.32\% & mdadx.com 0.31\% &  flashtalking.com 0.29\% \\ 
        exoclick.com 0.28\% & smartadserver.com 0.27\% & admission.net 0.20\% & ligatus.com 0.20\% & sitemeter.com 0.18\% \\ 
        atwola.com 0.17\% & teads.tv 0.17\% & adition.com 0.16\% & rfihub.com 0.13\% & atdmt.com 0.11\% \\ 
 
    \bottomrule
  \end{tabular}
\end{table*}

Table~\ref{tab:frequent_trackers} shows the list of frequent trackers overall and in some of the sensitive category web pages.
For each third-party" domain, we show the percentage of total web pages (browsed by all users) where it was present.
We find that, irrespective of the categories of web pages, Google and Facebook domains as third-party domains are prevalent.

\begin{table*}[H]
\centering
\caption{Top-10 third-party domains present overall and on some sensitive category web pages}
\begin{tabular}{|cl|cl|cl|}
\hline
\parbox[t]{2mm}{\multirow{10}{*}{\rotatebox[origin=c]{90}{\textbf{Overall}}}} & fbcdn.net 25.55\% 
& \parbox[t]{2mm}{\multirow{10}{*}{\rotatebox[origin=c]{90}{\textbf{Adult}}}} & google-analytics.com 47.06\%    
& \parbox[t]{2mm}{\multirow{10}{*}{\rotatebox[origin=c]{90}{\textbf{Religion}}}} & gstatic.com 44.80\%   \\
& doubleclick.net 20.44\% & & youtube.com  23.52\% && www.google.com 42.5\%\\
& google-analytics.com 17.77\% & & googleadservices.com 21.42\% && apis.google.com 42.30\%\\
& akamaihd.net 13.78\% & & doubleclick.net 19.59\% && google-analytics.com 11.53\%\\
& www.google.com 13.45\% & &  googlesyndication.com 17.16\% && amazonaws.com 7.69\%\\
& gstatic.com 12.62\% & & twitter.com 15.62\% && facebook.com 5.38\%\\
& accounts.google.com 9.74\% & & apis.google.com 13.29\% && youtube.com 4.23\%\\
& googleadservices.com 8.97\% & & www.google.com 12.22\% && ytimg.com 3.65\%\\
& googlesyndication.com 8.41\% & & googlevideo.com 11.51\% && xiti.com 3.27\%\\
& apis.google.com 8.38\% & & ytimg.com 9.82\% && doubleclick.net 3.07\%\\
\hline
\parbox[t]{2mm}{\multirow{10}{*}{\rotatebox[origin=c]{90}{\textbf{Health \& Fitness}}}} & google-analytics.com 24.37\% 
& \parbox[t]{2mm}{\multirow{10}{*}{\rotatebox[origin=c]{90}{\textbf{Military}}}} & gstatic.com 19.13\%    
& \parbox[t]{2mm}{\multirow{10}{*}{\rotatebox[origin=c]{90}{\textbf{Personal Finance}}}} & google-analytics.com 26.37\%   \\
& youtube.com 13.89\% & & apis.google.com 18.39\% && doubleclick.net 21.82\%\\
& maps.google.com 13.50\% & & plus.google.com 17.58\% && cloudfront.net 15.98\%\\
& fonts.net 13.19\% & & googleusercontent.com 17.25\% && googleadservices.com 14.23\%\\
& ytimg.com 13.12\% & &  googleadservices.com 17.18\% && scorecardresearch.com 14.00\%\\
& facebook.com 10.63\% & & www.google.com 16.37\% && googlesyndication.com 12.71\%\\
& amazonaws.com 10.09\% & & google-analytics.com 8.15\% && googletagservices.com 12.01\%\\
& maps.googleapis.com 8.30\% & & accounts.google.com 4.11\% && facebook.com 11.66\%\\
& doubleclick.net 6.75\% & & doubleclick.net 3.77\% && google.co.in 10.50\%\\
& facebook.net 6.05\% & & play.google.com 3.36\% && facebook.net 9.21\%\\
\hline
\end{tabular}
\label{tab:frequent_trackers}
\end{table*}

\end{document}